%% file: BobSimonLMCS.tex
\documentclass{LMCS} 
\def\doi{8(4:14)2012}
\lmcsheading%
{\doi}
{1--24}
{}
{}
{Jan.~25, 2011}
{Oct.~19, 2012}
{}   

\usepackage{amssymb} 
\usepackage{amsfonts}
\usepackage{amsmath}
\usepackage{latexsym}
\usepackage{epsfig}
\usepackage{verbatim}
\usepackage{units}
\usepackage{fix-cm}
\usepackage{stmaryrd}
\usepackage{xspace}
\usepackage{graphicx}
\usepackage{xy}
\xyoption{all}
\usepackage{color}
\usepackage{subfigure} 

\usepackage{enumerate}
\usepackage{hyperref}

\usepackage{tikz}
\usetikzlibrary[shapes.geometric]
\usepackage{tikz-cat}

\newcommand{\bit}{\begin{itemize}}
\newcommand{\eit}{\end{itemize}\par\noindent}
\newcommand{\ben}{\begin{enumerate}}
\newcommand{\een}{\end{enumerate}\par\noindent}
\newcommand{\beq}{\begin{equation}}
\newcommand{\eeq}{\end{equation}\par\noindent}
\newcommand{\beqa}{\begin{eqnarray*}}
\newcommand{\eeqa}{\end{eqnarray*}\par\noindent}
\newcommand{\beqn}{\begin{eqnarray}}
\newcommand{\eeqn}{\end{eqnarray}\par\noindent}

\def\C{{\bf C}}
\def\Cpure{{\bf C^{pure}}}
\def\D{{\bf Dens}}
\def\Dpure{{\bf Dens^{pure}}}

\def\II{{\rm I}}

\title[Environment and classical channels in CQM]{Environment and classical channels \\  in categorical quantum mechanics} 
\author[B.~Coecke]{Bob Coecke\rsuper a}	
\address{{\lsuper a}University of Oxford, Department of Computer Science, Quantum Group}	
\email{coecke@cs.ox.ac.uk}  

\author[S.~Perdrix]{Simon Perdrix\rsuper b}	
\address{{\lsuper b}CNRS, Laboratoire d'Informatique de Grenoble}
\email{simon.perdrix@imag.fr}  


\begin{document}

\maketitle

\input{pics.tex}

\begin{abstract}
We present a both simple and comprehensive graphical calculus  for 
quantum computing. In particular, we axiomatize the notion of an \emph{environment}, which together with the earlier introduced axiomatic notion of classical structure  enables us to define classical channels,  quantum measurements  and  classical control.  If we moreover adjoin the earlier introduced axiomatic notion of complementarity, we obtain sufficient structural power for  constructive representation and elegant correctness derivation of many quantum informatic protocols, including classically controlled quantum teleportation and dense coding, and quantum key distribution.
\end{abstract}

\section{Introduction}

Categorical quantum mechanics \cite{AC} provides a new perspective on quantum  information processing.   Particularly appealing  is the fact that the symmetric monoidal categorical language   comes with an intuitive  graphical calculus \cite{Penrose,JoyalStreet}.  This approach has meanwhile led to new results in quantum information and quantum foundations.     

Graph states, a key resource  for the measurement based quantum computational model, and translations thereof to the circuit model  were studied in  \cite{DuncanPerdrix,DuncanPerdrix2}.  There is a compositional framework for studying the structure of general multipartite quantum entanglement  \cite{CK}. Quantum theory as well as Spekkens' toy theories have been casted within a single mathematical framework, enabling  corresponding analysis of quantum non-locality in terms of `phase groups'  \cite{CES}.   There exists a no-cloning theorem for a very general class of theories \cite{AbrClone}.  

Meanwhile there also exists a software tool  with not only a graphical interface  but also a `graphical internal logic', \cite{quantomatic}, which (semi-)automates graphical reasoning.  It has been developed in a collaboration between Oxford, Edinburgh and since recently Google, and involved interpreting theories formulated in symmetric monoidal language in terms of `open graphs' \cite{Dixon:2010aa}. An important fragment of the graphical language admits a completeness theorem with respect to Hilbert spaces \cite{SelingerCompleteness}. 

Categorical quantum mechanics notions relevant for this paper are:
\ben[{\bf (A)}]
\item by endowing compact categories  \cite{KellyLaplaza} with an identity-on-objects  `dagger'-involution, one can  abstractly capture bipartite  states, map-state duality, bra-ket duality  and unitary operations  \cite{AC};
\item one can assign to any dagger compact category $\C$ of pure states and operations another dagger compact category $CPM(\C)$ of `mixed states' and `completely positive maps' \cite{Selinger};   $CPM(\C)$ can be axiomatized  as a dagger compact category with for every object $A$ a privileged `maximally mixed state' $\bot_A:\II\to A$ \cite{SelingerAxiom};
\item `classical structures' \cite{CPav,CPV}, which are certain kinds of commutative Frobenius algebras \cite{CarboniWalters}, enable one to handle classical data and control \cite{CPaqPav};
\item from an axiomatization of `complementarity'  (or `unbiasedness')  basic quantum logic gates can be constructed \cite{CD1,CD2}.
\een

\noindent The interaction of  these concepts has not been subjected to a detailed study yet.  Here, we distill the notion of \em environment \em out of {\bf (B)}, and by blending it in an appropriate manner with 
{\bf (C)} we define the notion of \em classical channel\em;  these  interact in a particularly nice manner with {\bf (D)} and 
together they provide a simple and elegant graphical calculus to represent and prove correctness of typical quantum computational protocols e.g.~teleportation (including classical control) \cite{BBC}, dense coding \cite{BW}, and QKD protocols \cite{BB,Ekert91}.  

Our main point is to show how with very little structural effort one straightforwardly reproduces these non-trivial quantum behaviors.   Furthermore, the simple explicit account that we obtain here on the classical-quantum interaction, which substantially simplifies the earlier work in this direction in \cite{CPaqPav},  will enable a fully comprehensive purely diagrammatic study of quantum informatic situations involving complex information flows between the classical and the quantum.  It moreover provides novel foundational insights on the nature of this interaction. 

Section \ref{sec:classicality} outlines how we conceive the classical-quantum distinction.  Section \ref{sec:background} recalls the notion of  \em classical structure\em,  and Section \ref{sec:compclassstruc} discusses complementary (or unbiasedness) thereof.  Section \ref{sec:environment} introduces the notion of \em environment\em, and Section \ref{sec:channel} combines environment and classical structure to form a \em classical channel\em, \em measurements \em   (\S\ref{subsec:meas})  and \em control operations \em  (\S\ref{subsec:cont}), and studies the role of \em complementarity \em 
 --\S\ref{sec:Densinterp} provides an explicit interpretation of the graphical language within Hilbert space quantum mechanics for the specific case of qubits.  In Section \ref{sec:examples} we derive basic quantum informatic protocols.   While we restrict ourselves to pure states and operations, our graphical framework  straightforwardly extends to mixed states and operations, as indicated in Section \ref{sec:Selinger}.

We assume that the reader is familiar with the basic concepts of quantum computation, such as states, operations, measurement and control.  The graphical language for symmetric monoidal categories is surveyed in \cite{SelingerSurvey}, and a tutorial tailored   towards applications in categorical quantum mechanics is \cite{CatsII}.
In this context, chapters \S2 and \S4 of \cite{CD2} are also useful.

\section{Classicality vs.~quantumness}\label{sec:classicality}

Let $\mathcal{ H}$ be a Hilbert space.  By a \em cloning map \em  one means an operation 
\[
U:\mathcal{ H}\otimes \mathcal{ H}\to \mathcal{ H}\otimes \mathcal{ H}
\]
which is such that for  all $|\psi\rangle\in \mathcal{ H}$ and some $|0\rangle\in \mathcal{ H}$ we have  
\[
U(|\psi\rangle\otimes|0\rangle)=|\psi\rangle\otimes|\psi\rangle\,.
\]
When fixing $|0\rangle$ within the argument we can instead consider
\[
\Delta:=U(-\otimes|0\rangle):\mathcal{ H}\to \mathcal{ H}\otimes \mathcal{ H}
\]
rather than $U$.    It is well-known that there exists no cloning map, by the so-called no-cloning theorem \cite{Dieks,WZ}. 
Now,  if an operation clones certain pure states, say the basis vectors $\{|i\rangle\}_i$, this does not imply that it clones mixtures of these too;  setting $\rho=\sum_i\rho_i |i\rangle\langle i|$ we have:
\[
\Delta\circ\rho\circ\Delta
=\Delta\circ\left(\sum_i\rho_i |i\rangle\langle i|\right)\circ\Delta
=\sum_i\rho_i \Delta\circ\left(|i\rangle\langle i|\right)\circ\Delta
=\sum_i\rho_i |ii\rangle\langle ii| \not=\rho\otimes\rho\,.
\]
However, what we do have is that  
\beq\label{eq:broadcasting1}
tr_1(\Delta\circ\rho\circ\Delta)=tr_2(\Delta\circ\rho\circ\Delta)=\rho\,,
\eeq
where $tr_1$ and $tr_2$ respectively trace out the first and the second system.  
For arbitrary completely positive maps   $\mathcal{ E}$ Eqs.~(\ref{eq:broadcasting1}) becomes:
\beq\label{eq:broadcasting2}
tr_1(\mathcal{ E}(\rho))=tr_2(\mathcal{ E}(\rho))=\rho\,.
\eeq
Universal validity of Eqs.~(\ref{eq:broadcasting2}) for a particular completely positive map $\mathcal{ E}$ acting on the entire space of all density operators has been referred to as \em broadcasting\em. 
But the \em no-broadcasting theorem \em \cite{Broadcast} states that  only mixed states which share a basis in which they all are diagonal (i.e.~mixed states that  can be   jointly  simulated by classical probability distributions) can be broadcast by the same completely positive map. 

From the above discussion it follows that broadcastability is a strictly weaker requirement than cloneability.  We have:
\begin{center}
\begin{tabular}{c|c|c|c|c|}
 & pure classical  & mixed classical &  pure quantum  & mixed quantum \\
\hline
broadcastable: &   {\tt YES}   &   \underline{\tt YES} &   {\tt NO}   &   {\tt NO}  \\
\hline
cloneable: &   {\tt yes}   &   \underline{\tt no} &   {\tt no}   &   {\tt no}  \\
\hline
\end{tabular} 
\end{center}
 where by classical we refer to a set of density operators that are diagonal in the same basis.  So (not non-cloneability but) non-broadcastability `identifies' quantum relative to classical, in that classical states, both pure and mixed, can always be broadcast by a single quantum operation, while this is not possible for quantum states that cannot be jointly simulated classically. Now, taking the contrapositive,    for us classicality will mean \em broadcastability\em.  

Equivalently, one can also conceive classicality as the result of \em decoherence \em \cite{Zurek}. Concretely, `total' decoherence is the completely positive map which erases all non-diagonal elements in the matrix representation of a given basis, that is,
\beq\label{eq:classex3}
\mathcal{ D}_{\{|i\rangle\}_i}::|ij\rangle\mapsto \delta_{ij}|ii\rangle\,,
\eeq
where $\delta_{ij}$ is the Kronecker delta.  Broadcasting and decoherence are indeed closely related: in the case of the first  `one copies into the environment', while in the case of the second `one  couples to the environment'. \em Decoherent \em then means invariance under this coupling.   Formally, decoherent density operators relative to a fixed $\mathcal{ D}_{\{|i\rangle\}_i}$  are exactly those collections of density operators that can be jointly broadcast; they all are diagonal in the basis  $\{|i\rangle\}_i$ and hence can be simulated by the classical probability distributions that make up the diagonals.

Summarizing the above: 
\begin{center}
\fbox{classical $:=$ broadcastable $\equiv$ decoherent} 
\end{center}
We will treat `classicality' as a `behavior' --i.e.~behaves as if it is classical in the above discussed sense-- rather than as the specification of the actual physical realization of a system.   An important point in this context, already realized in \cite{CPav,CPaqPav}, is that by taking quantum to be the `default behavior' within the mathematical universe of all operations, characterization of classical entities can be done in purely diagrammatic terms.  In the concrete Hilbert space realization, this means that one  only needs to rely on the \em multiplicative \em tensor product structure as a primitive connective, with no reference to the \em additive \em vector space structure.  One could refer to this as `classicization', in contrast to the standard notion of `quantization' where one starts with a classical theory and then freely adjoins the additive vector space structure. 

As an example, consider  a quantum measurement, which when applied to  a quantum system changes the state of that quantum system and produces classical data.  Since the resulting quantum state  is an eigenstate for that measurement, hence broadcastable, it behaves precisely in the same manner as the classical data does.   As a result, in the graphical calculus the classical data and the collapsed state won't be distinguishable once we omit explicit specification wether physically  they are either classical or quantum.  This ambiguity captures a \em feature \em of the particular manner in which quantum and classical data interact, namely, that the creation of classical data renders the quantum state in an eigenstate.  Of course, if we later apply a non-classical unitary to the resulting quantum state, then we reassert its proper  quantumness.  

Put in type-theoretic terms, there will be no such thing as  a fixed `classical type' and fixed `quantum type' in our representation, since we can abstract away over these  `implementation details' without altering the essential structure.  Of course, one can add those details in order to connect the graphical language to concrete physical protocols where the classical quantum distinction may be fundamental for the conceptual analysis, for example, in quantum teleportation it is  crucial that the classical communication can indeed be realized by purely classical finitary means.  In Section \ref{sec:examples} we give several examples of protocols that come with specification of what is classical and what is quantum, and then pass to the abstract diagrammatic calculus 
where forgetting the physical realization is essential to perform the computation.

\section{Classical structures}\label{sec:background}

In this paper we will work in the graphical representation of symmetric monoidal categories  \cite{JoyalStreet}.
 Mac Lane's strictification theorem \cite[p.257]{SML} allows us to take our symmetric monoidal categories to be strict,  that is:
\[
( A\otimes B)\otimes C=A\otimes (B\otimes C)\qquad\mbox{and}\qquad A\otimes \II=A=\II\otimes A\,.
\]
Morphisms $f:A_1\ldots A_n\to B_1\ldots B_m$, which we interpret as processes are respectively represented as boxes where the input wires represent the objects $A_1\ldots A_n$ and the output wires represent the objects $B_1\ldots B_m$:
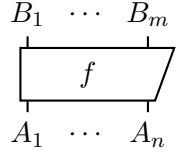
\[
\begin{tikzpicture}[cat,scale=0.8]
\node (in1) at (0,-1) {$A_1$};
\node (inn) at (2,-1) {$A_n$};
\node (in) at (1,-1) {$\cdots$};
\node (out1) at (0,1) {$B_1$};
\node (outn) at (2,1) {$B_m$};
\node (out) at (1,1) {$\cdots$};
\node [morph] (f) at (1,0) {$\quad f \quad$};
\draw[thick] (in1) -- (0,-0.45);
\draw[thick] (inn) -- (2,-0.45);
\draw[thick] (out1) -- (0,0.45);
\draw[thick] (outn) -- (2,0.45);
\end{tikzpicture}
\]
 Other shapes may be used to emphasize extra structure.
\em Elements \em $s:\II\to A_1\ldots A_n$, which in the graphical representation have no inputs, are interpret as `states', and \em co-elements \em $e: B_1\ldots B_m\to\II$ with no outputs are interpreted as `effects'.  In standard quantum notation they would be kets $|\psi\rangle$ and bras $\langle \psi|$ respectively. Composition and tensoring are respectively represented as:
\[
\begin{tikzpicture}[cat,scale=0.7]
\node (inf) at (1,-1) {};
\node (outf) at (1,1) {};
\node [morph] (f) at (1,0) {$g\circ f$};
\draw[thick] (inf) -- (f);
\draw[thick] (f) -- (outf);
\end{tikzpicture}\ \ \ 
\centering{=}\ \ \ \begin{tikzpicture}[cat,scale=0.7]
\node (in) at (1,-1) {};
\node (out) at (1,2.2) {};
\node [morph] (f) at (1,0) {$\hspace{0.05cm}f$};
\node [morph] (g) at (1,1.2) {$g$};
\draw[thick] (in) -- (f);
\draw[thick] (f) -- (g);
\draw[thick] (g) -- (out);
\end{tikzpicture}\quad\quad\quad\quad\quad\quad
\begin{tikzpicture}[cat,scale=0.7]
\node (inf) at (1,-1) {};
\node (outf) at (1,1) {};
\node [morph] (f) at (1,0) {$f\otimes g$};
\draw[thick] (inf) -- (f);
\draw[thick] (f) -- (outf);
\end{tikzpicture}\ \ \ 
\centering{=}\ \ \ \begin{tikzpicture}[cat,scale=0.7]
\node (inf) at (1,-1) {};
\node (outf) at (1,1) {};
\node (ing) at (2.5,-1) {};
\node (outg) at (2.5,1) {};
\node [morph] (f) at (1,0) {$\hspace{0.05cm}f$};
\node [morph] (g) at (2.5,0) {$g$};
\draw[thick] (inf) -- (f);
\draw[thick] (f) -- (outf);
\draw[thick] (g) -- (outg);
\draw[thick] (g) -- (ing);
\end{tikzpicture}
\]

\begin{thm}{\bf\cite{JoyalStreet,SelingerSurvey}}\label{thm:JoyalStreet}
An equation follows from the axioms of symmetric monoidal categories if and only if it can be derived in the graphical language via diagram isomorphisms.
\end{thm}

A \em dagger functor \em on a symmetric monoidal category \cite{Selinger} is an identity-on-objects contravariant involutive strict monoidal functor.  It is graphically represented by flipping pictures upside-down, for instance:
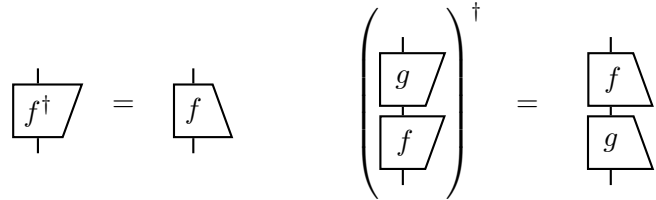
\[
\begin{tikzpicture}[cat,scale=0.7]
\node (inf) at (1,-1) {};
\node (outf) at (1,1) {};
\node [morph] (f) at (1,0) {$f^\dagger$};
\draw[thick] (inf) -- (f);
\draw[thick] (f) -- (outf);
\end{tikzpicture}\ \ \ 
\centering{=}\ \ \ \
\begin{tikzpicture}[cat,scale=0.7]
\node (inf) at (1,-1) {};
\node (outf) at (1,1) {};
\node [dagger morph] (f) at (1,0) {$f$};
\draw[thick] (inf) -- (f);
\draw[thick] (f) -- (outf);
\end{tikzpicture}\qquad\qquad
\left(\begin{tikzpicture}[cat,scale=0.7]
\node (in) at (1,-1) {};
\node (out) at (1,2.2) {};
\node [morph] (f) at (1,0) {$\hspace{0.05cm}f$};
\node [morph] (g) at (1,1.2) {$g$};
\draw[thick] (in) -- (f);
\draw[thick] (f) -- (g);
\draw[thick] (g) -- (out);
\end{tikzpicture}\right)^\dagger \ \ \ 
\centering{=}\ \ \  \ \
\begin{tikzpicture}[cat,scale=0.7]
\node (in) at (1,-1) {};
\node (out) at (1,2.2) {};
\node [dagger morph] (f) at (1,0) {$g$};
\node [dagger morph] (g) at (1,1.2) {$\hspace{0.05cm}f$};
\draw[thick] (in) -- (f);
\draw[thick] (f) -- (g);
\draw[thick] (g) -- (out);
\end{tikzpicture}
\]
Given such a dagger functor, a morphism $f:A\to B$ is an \em isometry \em if $f^\dagger\circ f=1_A$, and it is \em unitary \em of both $f$ and $f^\dagger$ are isometries. 

A \em dagger compact category \em \cite{AC2} is a dagger symmetric monoidal category in which each object $A$ comes with two morphisms $\eta_A: \II\to A^*\otimes A$ and $\epsilon_A:A\otimes A^*\to\II$ which satisfy certain equations.  In this paper we take all our objects to be \em self-dual\em, that is, $A=A^*$.\footnote{A detailed study of the coherences for this situation is in \cite{SelingerSelfDual}.} Graphically, we represent $\eta_A$ as:
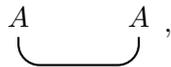
\[
\begin{tikzpicture}[cat,scale=0.8]
\node (in1) at (-1,0.8) {$A$};
\node (in2) at (1,0.8) {$A$};
\draw[thick,rounded corners=8pt] (in1) -- (-1,0) -- (1,0) -- (in2);
\end{tikzpicture}\,,
\]
we take $\epsilon_A$ to be its dagger, and the equations that govern $\eta_A$ and $\epsilon_A$ are:
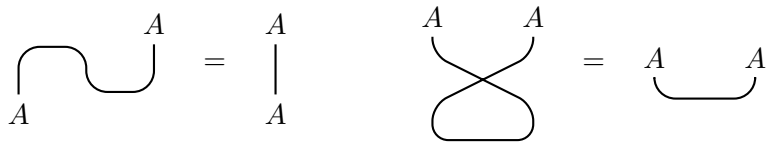
\[
\begin{tikzpicture}[cat,scale=0.6]
\node (in) at (0,0) {$A$};
\node (out) at (3,2) {$A$};
\draw[thick,rounded corners=8pt] (in) -- (0,1.5) -- (1.5,1.5) -- (1.5,0.5) -- (3,0.5) -- (out);
\end{tikzpicture}\ \ \ 
\centering{=}\ \ \ 
\begin{tikzpicture}[cat,scale=0.6]
\node (in) at (0,0) {$A$};
\node (out) at (0,2) {$A$};
\draw[thick] (in) -- (out) ;
\end{tikzpicture}\qquad\qquad
\begin{tikzpicture}[cat,scale=0.67]
\node (in1) at (-1,2.15) {$A$};
\node (in2) at (1,2.15) {$A$};
\draw[thick,rounded corners=6pt] (in1) -- (-1,1.4)-- (1,0.4) -- (1,-0.3) -- (-1,-0.3)--(-1,0.4)--(1,1.4)--(in2);
\end{tikzpicture}\ \ \ 
\centering{=}\ \ \ 
\begin{tikzpicture}[cat,scale=0.67]
\node (in1) at (-1,0.8) {$A$};
\node (in2) at (1,0.8) {$A$};
\draw[thick,rounded corners=8pt] (in1) -- (-1,0) -- (1,0) -- (in2);
\end{tikzpicture}
\]

\begin{rem}
The results in this paper can be extended to the case of non-self-dual compact structures, by relying on the results in \cite{CPaqPer}. This would, for example, be required when considering all three complementary measurements on a qubit. 
\end{rem}

\begin{thm}{\bf\cite{KellyLaplaza,Selinger}} \label{thm:KellyLaplaza-Selinger}
An equation follows from the axioms of (dagger) compact categories if and only if it can be derived in the corresponding graphical language via isotopy.
\end{thm}

The key difference between \em diagram isomorphism \em as in Theorem \ref{thm:JoyalStreet} and \em isotopy \em  as in Theorem \ref{thm:KellyLaplaza-Selinger} is that diagram isomorphisms take specification of the boxes' inputs and outputs into account, while isotopy  abstracts away these roles.  Hence within the scope of  Theorem \ref{thm:KellyLaplaza-Selinger} only the topology of the diagrams matters.

By \em classical structures \em \cite{CPav} we mean internal commutative special dagger Frobenius algebras in a dagger compact category  for which we also require `compatibility with the compact structure' (see below).   We won't give an explicit definition here, but will rely on a remarkable normal form result that holds for morphisms build from this structure, namely,  any morphism 
\[
\Xi_n^m: \underbrace{A\otimes \ldots \otimes A}_n\to \underbrace{A\otimes \ldots \otimes A}_m
\]
obtained by composing and tensoring  the structural morphisms of a classical structure and the symmetric monoidal structure, and of which  the diagrammatic representation is connected, only depends on $n$ and $m$ \cite{Lack}. Graphically  we  represent this unique morphism  as an  \em $n+m$-legged spider\em\,:
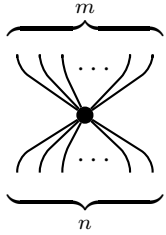
\[
\underbrace{\overbrace{\begin{tikzpicture}[cat,scale=0.6]
\node (in1) at (-0.5,-1) {};
\node (in2) at (0,-1) {};
\node (in3) at (0.5,-1) {};
\node (innm1) at (2,-1) {};
\node (inn) at (2.5,-1) {};
\node (in) at (1.25,-0.5) {$\cdots$};
\node (out1) at (-0.5,2) {};
\node (out2) at (0,2) {};
\node (out3) at (0.5,2) {};
\node (outn) at (2.5,2) {};
\node (outnm1) at (2,2) {};
\node (out) at (1.25,1.5) {$\cdots$};
\node [black vertex] (dot) at (1,0.5) {};
\draw[thick,rounded corners=5pt] (in1) -- (-0.5,-0.45) -- (dot) ;
\draw[thick,rounded corners=5pt] (in2) -- (0,-0.45) -- (dot);
\draw[thick,rounded corners=5pt] (in3) -- (0.5,-0.45) -- (dot) ;
\draw[thick,rounded corners=5pt] (innm1) -- (2,-0.45) -- (dot) ;
\draw[thick,rounded corners=5pt] (inn) -- (2.5,-0.45)-- (dot) ;
\draw[thick,rounded corners=5pt] (out1) -- (-0.5,1.55)-- (dot) ; 
\draw[thick,rounded corners=5pt] (out2) -- (0,1.55)-- (dot) ;
\draw[thick,rounded corners=5pt] (out3) -- (0.5,1.55)-- (dot) ;
\draw[thick,rounded corners=5pt] (outnm1) -- (2,1.55)-- (dot) ;
\draw[thick,rounded corners=5pt] (outn) -- (2.5,1.55)-- (dot) ;
\end{tikzpicture}}^{\scriptstyle{m}}}_{\scriptstyle{n}}
\]
From the axioms of classical structures it follows that these spiders are invariant when one exchanges the roles of front-legs and back-legs,  when one swaps two legs of  either of these,  and that the $(1+1)$-legged spider is the identity,  that is, 
\beq\label{eq:swapleggsdaggerleggs}
(\Xi_n^m)^\dagger=\Xi_m^n\ 
\quad\
 \Xi_n^m\circ (1_{A^{\otimes(k)}} \otimes\sigma_{A,A} \otimes 1_{A^{\otimes(n-k-2)}})= \Xi_n^m\
\quad\ \Xi_1^1=1_A\,,
\eeq
where $\sigma_{A,A}:A\otimes A\to A\otimes A$ is the swap map, and, last but not least,  that spiders which `share' legs fuse together, i.e.~spiders compose as follows:
\beq\label{eq:spidercomposition}
\underbrace{\overbrace{\begin{tikzpicture}[cat,scale=0.6]
\node (in1) at (-1,-1) {};
\node (in2) at (0.5,-1) {};
\node (innm1) at (1,-1) {};
\node (inn) at (3,-1) {};
\node (in) at (-0.25,-0.5) {$\cdots$};
\node (in) at (2,-0.5) {$\cdots$};
\node (m) at (1,0.5) {$\cdots$};
\node (out1) at (-1,2) {};
\node (out2) at (1,2) {};
\node (outn) at (3,2) {};
\node (outnm1) at (1.5,2) {};
\node (out) at (0,1.5) {$\cdots$};
\node (out) at (2.25,1.5) {$\cdots$};
\node [black vertex] (dot1) at (0,1) {};
\node [black vertex] (dot2) at (2,0) {};
\draw[thick,rounded corners=5pt](in1) -- (-1,-0.45)-- (dot1);
\draw[thick,rounded corners=5pt] (in2) -- (0.5,-0.45)-- (dot1);
\draw[thick,rounded corners=5pt] (innm1) -- (1,-0.45)-- (dot2);
\draw[thick,rounded corners=5pt] (inn) -- (3,-0.45)-- (dot2);
\draw[thick,rounded corners=5pt] (out1) -- (-1,1.55)-- (dot1);
\draw[thick,rounded corners=5pt] (out2) -- (1,1.55)-- (dot1);
\draw[thick,rounded corners=5pt] (outnm1) -- (1.5,1.55)-- (dot2);
\draw[thick,rounded corners=5pt] (outn) -- (3,1.55)-- (dot2);
\draw[thick,rounded corners=5pt] (dot1) -- (0.6,0.3)-- (dot2);
\draw[thick,rounded corners=5pt] (dot1) -- (1.4,0.7)-- (dot2);
\end{tikzpicture}}^{\scriptstyle{m}}}_{\scriptstyle{n}}~~~~\centering{=}~~~~
\underbrace{\overbrace{\begin{tikzpicture}[cat,scale=0.6]
\node (in1) at (-0.5,-1) {};
\node (in2) at (0,-1) {};
\node (in3) at (0.5,-1) {};
\node (innm1) at (2,-1) {};
\node (inn) at (2.5,-1) {};
\node (in) at (1.25,-0.5) {$\cdots$};
\node (out1) at (-0.5,2) {};
\node (out2) at (0,2) {};
\node (out3) at (0.5,2) {};
\node (outn) at (2.5,2) {};
\node (outnm1) at (2,2) {};
\node (out) at (1.25,1.5) {$\cdots$};
\node [black vertex] (dot) at (1,0.5) {};
\draw[thick,rounded corners=5pt] (in1) -- (-0.5,-0.45) -- (dot) ;
\draw[thick,rounded corners=5pt] (in2) -- (0,-0.45) -- (dot);
\draw[thick,rounded corners=5pt] (in3) -- (0.5,-0.45) -- (dot) ;
\draw[thick,rounded corners=5pt] (innm1) -- (2,-0.45) -- (dot) ;
\draw[thick,rounded corners=5pt] (inn) -- (2.5,-0.45)-- (dot) ;
\draw[thick,rounded corners=5pt] (out1) -- (-0.5,1.55)-- (dot) ;
\draw[thick,rounded corners=5pt] (out2) -- (0,1.55)-- (dot) ;
\draw[thick,rounded corners=5pt] (out3) -- (0.5,1.55)-- (dot) ;
\draw[thick,rounded corners=5pt] (outnm1) -- (2,1.55)-- (dot) ;
\draw[thick,rounded corners=5pt] (outn) -- (2.5,1.55)-- (dot) ;
\end{tikzpicture}}^{\scriptstyle{m}}}_{\scriptstyle{n}}
\eeq
Conversely, the axioms of an internal  commutative special dagger Frobenius algebra all follow from  Eqs.~(\ref{eq:swapleggsdaggerleggs}) and (\ref{eq:spidercomposition}).

By \em compatibility \em of the classical structure with a given  dagger  compact structure we mean that for a spider on $A$ we have $\eta_A=\Xi_2^0$, and consequently that $\epsilon_A=\Xi_0^2$.  In graphical terms, that is:
\[
\begin{tikzpicture}[cat,scale=0.6]
\node (in1) at (-1,0.8) {};
\node (in2) at (1,0.8) {};
\node[black vertex]  (dot) at (0,0) {};
\draw[thick,rounded corners=5pt]  (dot) -- (-1,0.4) -- (in1);
\draw[thick,rounded corners=5pt]  (dot) -- (1,0.4) -- (in2);
\end{tikzpicture}
\ \ \ \raisebox{-1.5mm}{\centering{=}}\ \ \ 
\begin{tikzpicture}[cat,scale=0.6]
\node (in1) at (-1,0.8) {};
\node (in2) at (1,0.8) {};
\draw[thick,rounded corners=8pt]  (in1) -- (-1,-0) -- (1,-0) -- (in2);
\end{tikzpicture}\qquad\qquad
\raisebox{-1.5mm}{\begin{tikzpicture}[cat,scale=0.6]
\node (in1) at (-1,-0.8) {};
\node (in2) at (1,-0.8) {};
\node[black vertex]  (dot) at (0,0) {};
\draw[thick,rounded corners=5pt]  (dot) -- (-1,-0.4) -- (in1);
\draw[thick,rounded corners=5pt]  (dot) -- (1,-0.4) -- (in2);
\end{tikzpicture}
\ \ \ \raisebox{-0.5mm}{\centering{=}}\ \ \ 
\begin{tikzpicture}[cat,scale=0.6]
\node (in1) at (-1,-0.8) {};
\node (in2) at (1,-0.8) {};
\draw[thick,rounded corners=8pt]  (in1) -- (-1,0) -- (1,0) -- (in2);
\end{tikzpicture}}
\]
Indeed, for spiders $\Xi_2^0$ and $\Xi_0^2$ we always have:
\[
\begin{tikzpicture}[cat,scale=0.6]
\node (in1) at (-1,-2) {};
\node (out1) at (3, 0.4) {};
\node[black vertex]  (dot) at (0,0) {};
\node[black vertex]  (dot2) at (2,-1.6) {};
\draw[thick,rounded corners=5pt]  (dot) -- (-1,-0.4) -- (in1);
\draw[thick,rounded corners=5pt]  (dot) -- (1,-0.4) -- (1,-1.2)--(dot2);
\draw[thick,rounded corners=5pt]  (dot2) -- (3,-1.2) -- (out1);
\end{tikzpicture}\ \ \ \raisebox{-0.5mm}{\centering{=}}\ \ \ 
\begin{tikzpicture}[cat,scale=0.6]
\node (in) at (0,-2) {};
\node[black vertex]  (dot) at (0,-0.8) {};
\node (out) at (0,0.4) {};
\draw[thick] (in) -- (out) ;
\end{tikzpicture}\ \ \ \raisebox{-0.5mm}{\centering{=}}\ \ \ 
\begin{tikzpicture}[cat,scale=0.6]
\node (in) at (0,-2) {};
\node (out) at (0,0.4) {};
\draw[thick] (in) -- (out) ;
\end{tikzpicture}
\]
and
\[
\begin{tikzpicture}[cat,scale=0.6]
\node (in1) at (-1,2.15) {};
\node (in2) at (1,2.15) {};
\node[black vertex]  (dot) at (0,-0.8) {};
\draw[thick,rounded corners=6pt] (in1) -- (-1,1.4)-- (1,0.4) -- (1,-0.4) -- (dot);
\draw[thick,rounded corners=6pt] (dot) -- (-1,-0.4)--(-1,0.4)--(1,1.4)--(in2);
\end{tikzpicture}\ \ \ 
\centering{=}\ \ \ 
\begin{tikzpicture}[cat,scale=0.6]
\node (in1) at (-1,0.8) {};
\node (in2) at (1,0.8) {};
\node[black vertex]  (dot) at (0,0) {};
\draw[thick,rounded corners=5pt]  (dot) -- (-1,0.4) -- (in1);
\draw[thick,rounded corners=5pt]  (dot) -- (1,0.4) -- (in2);
\end{tikzpicture}
\]
that is, they form a dagger compact structure.

In the graphical language we will depict elements (i.e.~`boxes without inputs') by triangles. 
By a  \em pure classical element \em for a particular classical structure we mean an element which satisfies:
\beq\label{eq:claselts}\begin{tikzpicture}[cat,scale=0.8]
\node (in1) at (-1,0.8) {};
\node (in2) at (1,0.8) {};
\node[black vertex]  (dot) at (0,0) {};
\node (e) at (0,-0.8) {$e$};
\draw[thick] (-0.5,-0.5) -- (0.5,-0.5) -- (0,-1.2) -- cycle;
\draw[thick] (0,-0.5) -- (dot);
\draw[thick,rounded corners=5pt]  (dot) -- (-1,0.4) -- (in1);
\draw[thick,rounded corners=5pt]  (dot) -- (1,0.4) -- (in2);
\end{tikzpicture}
\ \ \  \ \ \centering{=}\ \ \ \ \ 
\begin{tikzpicture}[cat,scale=0.8]
\node (e) at (0,-0.8) {$e$};
\draw[thick] (-0.5,-0.5) -- (0.5,-0.5) -- (0,-1.2) -- cycle;
\draw[thick] (0,-0.5) -- (0,0);
\node (e) at (1.5,-0.8) {$e$};
\draw[thick] (1,-0.5) -- (2,-0.5) -- (1.5,-1.2) -- cycle;
\draw[thick] (1.5,-0.5) -- (1.5,0);
\end{tikzpicture}
\eeq
i.e.~it is `copied'. Below, by $e$ we will only  denote such pure classical elements.

In the dagger compact category  ${\bf FHilb}$ which has finite dimensional Hilbert spaces as objects, linear maps as morphisms, the tensor product as the monoidal structure, and adjoints as the dagger, classical structures are in bijective correspondence orthonormal bases  via this concept of pure classical elements \cite{CPV}.   Concretely, 
the pure classical  elements are exactly the basis vectors, and conversely, given an orthonormal  basis $\{|i\rangle\}_i$, the corresponding spiders are the linear maps with as only non-zero action on the basis vectors:
\beq\label{eq:classex1}
\Xi_n^m::\underbrace{|i\ldots i\rangle}_n\mapsto \underbrace{|i\ldots i\rangle}_m\,,
\eeq
i.e.~arrays of identical basis vectors are mapped on arrays of identical basis vectors, and all other basis vectors are mapped on the zero vector.  Important particular examples are 
\beq\label{eq:classex2}
\Xi_2^1::|ij\rangle\mapsto\delta_{ij} |i\rangle \qquad\mbox{and}\qquad \Xi_0^1::1\mapsto\sum_i |i\rangle
\eeq
which define the \em multiplication \em and its \em unit \em of the corresponding Frobenius algebra; their adjoints 
define the corresponding comultiplication and its counit.  

\begin{rem}\label{rem:specifyspider}
In any dagger symmetric monoidal category, the multiplication and its unit suffice to specify a classical structure; one can then construct any other spider by composing $\Xi_2^1$, $\Xi_0^1$, $(\Xi_2^1)^\dagger$ and $(\Xi_0^1)^\dagger$ to obtain a morphism with the required number of inputs $n$ and outputs $m$, that is, the spider $\Xi_n^m$.
\end{rem}

\begin{rem}\label{rem:WP}
Physically relevant,  rather  than  ${\bf FHilb}$, is the category $W\!P({\bf FHilb})$  which is obtained by subjecting ${\bf FHilb}$  to the congruence which identifies those linear maps of the same type that are equal up to a complex phase, i.e.
\[
f\sim g\ \Leftrightarrow\ \exists \theta\in [0, 2\pi[: f=e^{i\theta}\cdot g\, .
\]
The reason is that vectors which are equal up to a complex phase represent the same state in quantum theory.  The precise connection between classical structures in ${\bf FHilb}$ and those in $W\!P({\bf FHilb})$ is studied in detail in \cite{CD2}; roughly put --since this suffices for all practical purposes-- classical structures are inherited.
\end{rem}

\section{Complementary classical structures}\label{sec:compclassstruc}

\begin{defi}
A  \emph{complementary}  endomorphism $H:A \to A$ for a classical structure is a (i) self-conjugate (ii) self-adjoint  (iii) unitary endomorphism, graphically,
\[
\begin{tikzpicture}[cat,scale=0.8]
\node (in1) at (-1,0.8){};
\node[h vertex] (h) at (-1,0.2) {};
\node (in2) at (0.5,0.8) {};
\draw[thick,rounded corners=8pt] (h) -- (in1);
\draw[thick,rounded corners=8pt] (h) -- (-1,-0.3) -- (0.5,-0.3) -- (in2);
\end{tikzpicture}\ \ \ \centering{=}\ \ \ 
\begin{tikzpicture}[cat,scale=0.8]
\node (in1) at (-1,0.8){};
\node[h vertex] (h) at (0.5,0.2) {};
\node (in2) at (0.5,0.8) {};
\draw[thick,rounded corners=8pt] (h) -- (in2);
\draw[thick,rounded corners=8pt] (h) -- (0.5,-0.3) -- (-1,-0.3) -- (in1);
\end{tikzpicture}
\qquad\qquad\qquad
\begin{tikzpicture}[cat,scale=0.8]
\node (in) at (-1,0.8){};
\node[h vertex] (h1) at (-1,0.2) {};
\node[h vertex] (h2) at (-1,-0.4) {};
\node (out) at (-1,-1) {};
\draw[thick,rounded corners=8pt] (h1) -- (in);
\draw[thick,rounded corners=8pt] (h1) -- (h2);
\draw[thick,rounded corners=8pt] (h2) -- (out);
\end{tikzpicture}\ \ \ \centering{=}\ \ \ 
\begin{tikzpicture}[cat,scale=0.8] 
\node (in) at (-1,0.5){};
\node (out) at (-1,-1) {};
\draw[thick,rounded corners=8pt] (in) -- (out);
\end{tikzpicture}
\]
where self-adjointness is encoded in the symmetry of the small box depicting $H$,
and, which `transforms a given classical structure into a complementary one', 
which --following \cite{CD2}-- graphically  depicts as:
\beq\label{eq:comp}
\begin{tikzpicture}[cat,scale=0.8]
\node (in) at (0,1) {};
\node [black vertex] (gdot) at (0,0.5) {};
\node[h vertex] (h1) at (-0.5,0) {};
\node[h vertex] (h2) at (0.5,0) {};
\draw[thick,rounded corners=2pt] (gdot) -- (0.5,0.3) -- (h2);
\draw[thick,rounded corners=2pt] (gdot) -- (-0.5,0.3) -- (h1);
\draw[thick] (gdot) -- (in);
\node (out) at (0,-1) {};
\node [black vertex] (gdot2) at (0,-0.5) {};
\draw[thick,rounded corners=2pt] (gdot2) -- (0.5,-0.3) -- (h2);
\draw[thick,rounded corners=2pt] (gdot2) -- (-0.5,-0.3) -- (h1);
\draw[thick] (gdot2) -- (out);
\end{tikzpicture}\ \ \ \centering{=}\ \ \ 
\begin{tikzpicture}[cat,scale=0.8]
\node (in) at (0,1) {};
\node [black vertex] (gdot) at (0,0.4) {};
\draw[thick] (gdot) -- (in);
\node (out) at (0,-1) {};
\node [black vertex] (gdot2) at (0,-0.4) {};
\draw[thick] (gdot2) -- (out);
\end{tikzpicture}\ \ .
\eeq
\end{defi}

\def\hadm{\raisebox{0mm}{$\frac 1{\sqrt 2}\left(\begin{array}{rr}
1 & 1\\
1 & -1
\end{array}\right)$}}

An example of such a  complementary   morphism is the familiar Hadamard matrix:
\[
\hadm\,,
\]
expressed in the basis defined by the classical structure.
Rather than axiomatizing a pair of complementary classical structures as in \cite{CD2},
here we axiomatize a morphism which transforms a classical structure in the other one. The defining equation in \cite{CD2}, which involves two classical structures, 
is obtained by postcomposing  both sides of Eq.~(\ref{eq:comp}) with $H$,  where  
\[
\tilde{\Xi}_2^1:= H \circ \Xi_2^1\circ (H\otimes H)\quad\qquad\mbox{ and } \qquad\quad
\tilde{\Xi}_0^1:= H\circ \Xi_0^1 
\] 
then respectively are the multiplication  and the unit  of the second classical structure.    Representing $(\Xi_2^1, \Xi_0^1)$ in green and $(\tilde{\Xi}_2^1, \tilde{\Xi}_0^1)$ in red,  we have:
\[
\begin{tikzpicture}[cat,scale=0.8]
\node (in) at (0,1) {};
\node [red vertex] (gdot) at (0,0.5) {};
\node [green vertex] (gdot2) at (0,-0.5) {};
\draw[thick,rounded corners=2pt] (gdot) -- (0.5,0.3) -- (0.5,-0.3) -- (gdot2);
\draw[thick,rounded corners=2pt] (gdot) -- (-0.5,0.3) --(-0.5,-0.3) -- (gdot2);
\draw[thick] (gdot) -- (in);
\node (out) at (0,-1) {};
\draw[thick] (gdot2) -- (out);
\end{tikzpicture}\ \ \ \centering{=}\ \ \ 
\begin{tikzpicture}[cat,scale=0.8]
\node (in) at (0,1.7) {};
\node[h vertex] (h0) at (0,1) {};
\node [black vertex] (gdot) at (0,0.5) {};
\node[h vertex] (h1) at (-0.5,0) {};
\node[h vertex] (h2) at (0.5,0) {};
\draw[thick,rounded corners=2pt] (gdot) -- (0.5,0.3) -- (h2);
\draw[thick,rounded corners=2pt] (gdot) -- (-0.5,0.3) -- (h1);
\draw[thick] (gdot) -- (h0)-- (in);
\node (out) at (0,-1) {};
\node [black vertex] (gdot2) at (0,-0.5) {};
\draw[thick,rounded corners=2pt] (gdot2) -- (0.5,-0.3) -- (h2);
\draw[thick,rounded corners=2pt] (gdot2) -- (-0.5,-0.3) -- (h1);
\draw[thick] (gdot2) -- (out);
\draw[dashed,rounded corners=4pt] (0,1.3) -- (-0.2,1.3)-- (-0.8,0.5)--(-0.8,-0.3) --(0.8,-0.3) -- (0.8,0.5) -- (0.2,1.3)--(0,1.3);
\end{tikzpicture}\ \ \ \centering{\overset{(\ref{eq:comp})} =}\ \ \ 
\begin{tikzpicture}[cat,scale=0.8]
\node (in) at (0,1.7) {};
\node[h vertex] (h0) at (0,1) {};
\node [black vertex] (gdot) at (0,0.4) {};
\draw[thick] (gdot) -- (h0)--(in);
\node (out) at (0,-1) {};
\node [black vertex] (gdot2) at (0,-0.4) {};
\draw[thick] (gdot2) -- (out);
\draw[dashed,rounded corners=4pt] (0,1.3) -- (-0.4,1.3)-- (-0.4,0.1)-- (0.4,0.1) -- (0.4,1.3)--(0,1.3);
\end{tikzpicture}
\ \ \ \centering{=}\ \ \ 
\begin{tikzpicture}[cat,scale=0.8]
\node (in) at (0,1) {};
\node [red vertex] (gdot) at (0,0.4) {};
\draw[thick] (gdot) --(in);
\node (out) at (0,-1) {};
\node [green vertex] (gdot2) at (0,-0.4) {};
\draw[thick] (gdot2) -- (out);
\end{tikzpicture}
\]
so we recover the characterization of complementarity as in \cite{CD2}\S8.

\begin{rem}
 Since $H$ is self-transposed we can set
\[
\begin{tikzpicture}[cat,scale=0.8]
\node (in1) at (-0.5,1) {};
\node (in2) at (0.5,1) {};
\node[h vertex] (h) at (0,0) {};
\node [black vertex] (gdot1) at (-0.5,0) {};
\node [black vertex] (gdot2) at (0.5,0) {};
\node (out1) at (-0.5,-1) {};
\node (out2) at (0.5,-1) {};
\draw[thick,rounded corners=2pt] (in1)--(gdot1) -- (out1);
\draw[thick,rounded corners=2pt] (in2)--(gdot2) -- (out2);
\draw[thick] (gdot1) -- (h)-- (gdot2);
\end{tikzpicture}\ \ \ \centering{:=}\ \ \ 
\begin{tikzpicture}[cat,scale=0.8]
\node (in1) at (-0.5,1) {};
\node (in2) at (0.5,1) {};
\node[h vertex] (h) at (0,0) {};
\node [black vertex] (gdot1) at (-0.5,-0.5) {};
\node [black vertex] (gdot2) at (0.5,0.5) {};
\node (out1) at (-0.5,-1) {};
\node (out2) at (0.5,-1) {};
\draw[thick,rounded corners=2pt] (in1)--(gdot1) -- (out1);
\draw[thick,rounded corners=2pt] (in2)--(gdot2) -- (out2);
\draw[thick,rounded corners=2pt] (gdot1) --(0,-0.3) -- (h)-- (0,0.3) -- (gdot2);
\end{tikzpicture}\ \ \ \centering{=}\ \ \ 
\begin{tikzpicture}[cat,scale=0.8]
\node (in1) at (-0.5,1) {};
\node (in2) at (0.5,1) {};
\node[h vertex] (h) at (0,0) {};
\node [black vertex] (gdot1) at (-0.5,0.5) {};
\node [black vertex] (gdot2) at (0.5,-0.5) {};
\node (out1) at (-0.5,-1) {};
\node (out2) at (0.5,-1) {};
\draw[thick,rounded corners=2pt] (in1)--(gdot1) -- (out1);
\draw[thick,rounded corners=2pt] (in2)--(gdot2) -- (out2);
\draw[thick,rounded corners=2pt] (gdot1) --(0,0.3) -- (h)-- (0,-0.3) -- (gdot2);
\end{tikzpicture}
\]
By Eq.~(\ref{eq:comp}) it then for example follows that:
\[
\begin{tikzpicture}[cat,scale=0.8]
\node (in1) at (-0.8,1) {};
\node (in2) at (0.8,1) {};
\node[h vertex] (h1) at (0,0.3) {};
\node[h vertex] (h2) at (0,-0.3) {};
\node [black vertex] (gdot1) at (-0.8,0) {};
\node [black vertex] (gdot2) at (0.8,0) {};
\node (out1) at (-0.8,-1) {};
\node (out2) at (0.8,-1) {};
\draw[thick,rounded corners=2pt] (in1)--(gdot1) -- (out1);
\draw[thick,rounded corners=2pt] (in2)--(gdot2) -- (out2);
\draw[thick,rounded corners=2pt] (gdot1) -- (-0.4,0.3) -- (h1)--(0.4,0.3) --  (gdot2);
\draw[thick,rounded corners=2pt] (gdot1) -- (-0.4,-0.3) -- (h2)--(0.4,-0.3) --  (gdot2);
\end{tikzpicture}\ \ \ \centering{=}\ \ \ 
\begin{tikzpicture}[cat,scale=0.8]
\node (in1) at (-0.8,1.5) {};
\node (in2) at (0.8,1.5) {};
\node[h vertex] (h1) at (0.3,0) {};
\node[h vertex] (h2) at (-0.3,0) {};
\node [black vertex] (gdot1) at (-0.8,-1) {};
\node [black vertex] (gdot2) at (0.8,1) {};
\node [black vertex] (gdot3) at (0,-0.5) {};
\node [black vertex] (gdot4) at (0,0.5) {};
\node (out1) at (-0.8,-1.5) {};
\node (out2) at (0.8,-1.5) {};
\draw[thick,rounded corners=2pt] (in1)--(gdot1) -- (out1);
\draw[thick,rounded corners=2pt] (in2)--(gdot2) -- (out2);
\draw[thick,rounded corners=2pt] (gdot1) --(0,-0.8) -- (gdot3);
\draw[thick,rounded corners=2pt] (gdot4)-- (0,0.8) -- (gdot2);
\draw[thick,rounded corners=2pt] (gdot3)-- (-0.3,-0.3) -- (h2) -- (-0.3,0.3) -- (gdot4);
\draw[thick,rounded corners=2pt] (gdot3)-- (0.3,-0.3) -- (h1) -- (0.3,0.3) -- (gdot4);
\end{tikzpicture}\ \ \ \centering{=}\ \ \ 
\begin{tikzpicture}[cat,scale=0.8]
\node (in1) at (-0.5,1.5) {};
\node (in2) at (0.5,1.5) {};
\node [black vertex] (gdot1) at (-0.5,-1) {};
\node [black vertex] (gdot2) at (0.5,1) {};
\node [black vertex] (gdot3) at (0,-0.5) {};
\node [black vertex] (gdot4) at (0,0.5) {};
\node (out1) at (-0.5,-1.5) {};
\node (out2) at (0.5,-1.5) {};
\draw[thick,rounded corners=2pt] (in1)--(gdot1) -- (out1);
\draw[thick,rounded corners=2pt] (in2)--(gdot2) -- (out2);
\draw[thick,rounded corners=2pt] (gdot1) --(0,-0.8) -- (gdot3);
\draw[thick,rounded corners=2pt] (gdot4)-- (0,0.8) -- (gdot2);
\end{tikzpicture}~~~\centering{=}~~~
\begin{tikzpicture}[cat,scale=0.8]
\node (in1) at (-0.5,1) {};
\node (in2) at (0.5,1) {};
\node (out1) at (-0.5,-1) {};
\node (out2) at (0.5,-1) {};
\draw[thick,rounded corners=2pt] (in1) -- (out1);
\draw[thick,rounded corners=2pt] (in2) -- (out2);
\end{tikzpicture}
\]
We will use this `topological trick' throughout this paper.
\end{rem}

\section{Environment}\label{sec:environment}

We will consider two dagger compact categories, denoted $\Cpure$ and ${\bf C}$ respectively, and we assume that 
$\Cpure$ is a subcategory of  ${\bf C}$ which inherits  symmetric monoidal structure as well as dagger compact structure, and that $|{\bf C}|=|\Cpure|$.

\begin{defi}\label{def:environment}  
An \em environment structure \em for $(\Cpure, {\bf C})$  consists of a designated co-element $\top_A:A\to\II$ for each object $A\in|\C|$, which we   depicted as:
\[
\begin{tikzpicture}[cat,scale=0.8]
\node (in) at (0,-1) {$A$};
\ground{gr}{0,0}
\draw[thick] (in) -- (gr);
\end{tikzpicture}\ \ ,
\]
and that  for all $A, B\in |\C|$ and all $ f, g\in \Cpure(A,B)$ we have
\beq\label{eq:environment2}
\begin{tikzpicture}[cat,scale=0.8]
\node (in) at (1,-1) {$A$};
\node [morph] (f) at (1,0) {$f$};
\node [dagger morph] (fdag) at (1,1) {$f$};
\node (out) at (1,2) {$A$};
\draw[thick] (in) -- (f);
\draw[thick] (f) -- (fdag);
\draw[thick] (out) -- (fdag);
\end{tikzpicture}
\ \ \  \
\centering{=}\ \ \ \
\begin{tikzpicture}[cat,scale=0.8]
\node (in) at (1,-1) {$A$};
\node [morph] (g) at (1,0) {$g$};
\node [dagger morph] (gdag) at (1,1) {$g$};
\node (out) at (1,2) {$A$};
\draw[thick] (in) -- (g);
\draw[thick] (g) -- (gdag);
\draw[thick] (out) -- (fdag);
\end{tikzpicture}
\ \ \ \
\centering{\iff}\ \ \ \
\begin{tikzpicture}[cat,scale=0.8]
\node (in) at (1,-1) {$A$};
\node [morph] (f) at (1,0) {$f$};
\ground{grf}{1,0.8}
\draw[thick] (in) -- (f);
\draw[thick] (f) -- (grf);
\end{tikzpicture}
\ \ \ \
\centering{=}\ \ \ \
\begin{tikzpicture}[cat,scale=0.8]
\node (in) at (1,-1) {$A$};
\node [morph] (g) at (1,0) {$g$};
\ground{grd}{1,0.8}
\draw[thick] (in) -- (g);
\draw[thick] (g) -- (grd);
\end{tikzpicture}
\eeq
\beq\label{eq:environment1}
\begin{tikzpicture}[cat,scale=0.8]
\node (in) at (0,-1) {$A\otimes B$};
\ground{gr}{0,0}
\draw[thick] (in) -- (gr);
\end{tikzpicture}\ \ \ \
\centering{=}\ \ \ \
\begin{tikzpicture}[cat,scale=0.8]
\node (in) at (0,-1) {$A$};
\ground{gr}{0,0}
\draw[thick] (in) -- (gr);
\end{tikzpicture}\ \ \ 
\begin{tikzpicture}[cat,scale=0.8]
\node (in) at (0,-1) {$B$};
\ground{gr}{0,0}
\draw[thick] (in) -- (gr);
\end{tikzpicture}
\qquad\qquad\qquad
\begin{tikzpicture}[cat,scale=0.8]
\node (in) at (0,1) {$A$};
\ground{gr}{1,0.6}
\draw[thick,rounded corners=8pt] (gr) -- (1,0) -- (0,0) -- (in);
\end{tikzpicture}\ \ \ \
\centering{=}\ \ \ \
\begin{tikzpicture}[cat,scale=0.8]
\node (in) at (1,1) {$A$};
\maxmixed{gr}{1,0.3}
\draw[thick] (in) -- (gr);
\end{tikzpicture}\ .
\eeq
\end{defi}

We will sometimes abbreviate environment structure to environment.

\begin{rem}
Eqs.~(\ref{eq:environment2},\ref{eq:environment1}) had  already been introduced in \cite{SelingerAxiom}, as part of an axiomatization of mixed states and completely positive maps, but it was never considered in relation to classicality, measurements, and complementarity thereof. 
\end{rem}

\begin{exa}
Let $\D$ be the category with  ${\mathbb C}^{n\times n}$ for 
$n\in \mathbb N$ as objects, and with completely positive maps $F: {\mathbb C}^{n\times n}\to{\mathbb C}^{m\times m}$ as morphisms. A morphism $f$ is  pure, that is, $f\in hom_\Dpure ({\mathbb C}^{n\times n},{\mathbb C}^{m\times m})$, if  there exists a linear map $L: {\mathbb C}^n \to {\mathbb C}^m$ such that  $f :: \rho\mapsto L\rho L^\dagger$.   Then , the usual trace
 \[
 \top_{\mathbb C^{n\times n}}:{\mathbb C}^{n\times n}\to {\mathbb C} :: \rho\mapsto tr(\rho)
 \]
 provides an environment structure for $(\Dpure, \D)$.   Indeed, for any $f::\rho\mapsto L\rho L^\dagger$ and $g::\rho\mapsto M\rho M^\dagger$, equation \ref{eq:environment2} is satisfied: 
\begin{eqnarray*}
f^\dagger \circ f = g^\dagger \circ g &\iff &\forall \rho: ~ L^\dagger (L\rho L^\dagger) L = M^\dagger (M\rho M^\dagger) M\\&\iff&
L^\dagger L=M^\dagger M \\&\iff &\forall \rho:  
~ tr(L^\dagger L\rho)= tr(M^\dagger M\rho)\\
&\iff&\forall \rho: 
~ tr(L\rho L^\dagger)= tr(M\rho M^\dagger)\\ 
&\iff & \top\circ f = \top \circ g\ .
\end{eqnarray*}
 This example justifies the name `environment': tracing a system out in quantum theory is interpreted as this system being part of the environment.
\end{exa}

\begin{exa}\label{Ex:CPM}
Let ${\bf C}$ be any dagger compact category, let $CPM({\bf C})$ the category obtained by applying Selinger's CPM-construction  \cite{Selinger} --we explicitly present this construction in Section \ref{sec:Selinger}-- and let $W\!P({\bf C})$ be the category obtained by applying the `doubling construction' of \cite{deLL} which cancels out `abstract global phases' --cf.~Remark \ref{rem:WP}.  Then, the caps of the dagger compact structure provide an environment for $(W\!P({\bf C}), CPM({\bf C}))$ \cite{SelingerAxiom}.
When taking ${\bf C}$ to be ${\bf FHilb}$ then we recover the previous example, with: 
\[
\D=CPM({\bf FHilb}) \qquad\mbox{and}\qquad \Dpure=WP({\bf FHilb})\, .
\]
\end{exa}

Below we assume as given a pair $(\Cpure, {\bf C})$ with an environment.  We set 
\beq
\bot_A:=\top_A^\dagger: {\rm I}\to A.   
\eeq
An element $\psi:\II\to A$ is \em normalized \em if 
\beq
\top_A\circ \psi=1_\II\,.   
\eeq
Below all elements depicted as triangles are normalized, diagrammatically:  
\beq\label{eq:normalize}
\begin{tikzpicture}[cat,scale=0.8]
\ground{grd}{0,0}
\node (e) at (0,-0.6) {\small $\psi$};
\draw[thick] (-0.4,-0.4) -- (0.4,-0.4) -- (0,-0.9) -- cycle;
\draw[thick] (0,-0.4) -- (grd);
\end{tikzpicture}\ \ \ \
\centering{=}\ \ \ \
\eeq
where $1_I$ is graphically represented by an empty picture.

\begin{prop}
A morphism $f\in\Cpure$ is an isometry iff  $\top_B\circ f=\top_A$, and hence, it is unitary iff we moreover have that $f\circ \bot_A=\bot_B$.
\end{prop}

\section{Classical channels, measurements and classical control}\label{sec:channel}

\begin{defi}
Let $\Xi$ be a classical structure. The morphism:
\[
C_{\Xi}\ \ =\ \ \ \begin{tikzpicture}[cat,scale=0.8]
\node (in) at (0,1.2) {};
\node (out) at (0,-0.8) {};
\ground{gr}{0.7,0.7}
\node [black vertex] (gdot) at (0,0) {};
\draw[thick,,rounded corners=2pt] (gdot) -- (0.7,0.5) -- (gr);
\draw[thick] (gdot) -- (in);
\draw[thick] (gdot) -- (out);
\end{tikzpicture}
\]
is called the \em classical channel of type $\Xi$\em.
\end{defi}

In the light of the discussion in Section \ref{sec:classicality}, this picture can be interpreted as `copying into the environment', that is, `broadcasting', or in the decoherence view, `being coupled to the environment'.   

\begin{exa}
In $\D$, for the classical structure of Eqs.~(\ref{eq:classex2}),
we have 
\[
C_\Xi(\rho)=  tr(\Xi_1^2\circ\rho\circ\Xi_2^1)=\mathcal{ D}_{\{|i\rangle\}_i}\,
\]
where $\mathcal{ D}_{\{|i\rangle\}_i}$ was defined in Eq.~(\ref{eq:classex3}).  That is, a classical channel preserves the diagonal data relative to the basis that is specified by its basis structure.  It will however destroy the non-diagonal data.  In the light of the discussion of Section \ref{sec:classicality}, it is a `classicizing' operation, on the data it is applied to. 
\end{exa}


\begin{rem}
While physically all classical channels are of course the same, our classical channels in addition carry specification of how the classical data it transmits has been obtained, in terms of a dependency on the classical structure $\Xi$ which specifies a particular quantum measurement.  In the light of the fact that by default we take all systems to be quantum, this specification of the classical structure relative to which classical data is classical is indeed unavoidable.  It is for this reason that we choose to  axiomatize the  complementary  morphism --cf.~the discussion in Section \ref{sec:compclassstruc}-- which enables us to restrict ourselves  to a single classical structure.  
\end{rem}

The following proposition shows that a classical channel leaves its pure classical elements invariant, and that it is idempotent.  In fact,  we could define  more general \em classical elements \em $p:\II\to A$ as those that satisfy 
\beq
C_{\Xi}\circ p=p\,.   
\eeq
Physically,   this means that a classical channel `transmits' its classical elements. Equivalently, classical elements are invariant under decoherence.

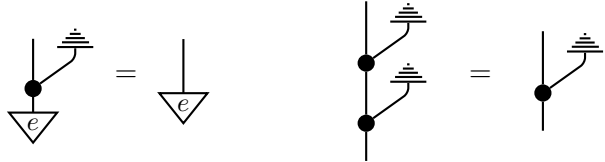
\begin{prop}\label{prop:idempotence}
\ \vspace{-5.5mm}\[
\begin{tikzpicture}[cat,scale=0.8]
\node (in) at (0,1) {};
\ground{gr1}{0.7,0.7}
\node (e) at (0,-0.6) {\small $e$};
\draw[thick] (-0.4,-0.4) -- (0.4,-0.4) -- (0,-0.9) -- cycle;
\draw[thick] (0,-0.4) -- (0,0);
\node [black vertex] (gdot1) at (0,0) {};
\draw[thick,rounded corners=2pt] (gdot1) -- (0.7,0.5)-- (gr1);
\draw[thick] (gdot1) -- (in);
\end{tikzpicture}\ \ 
\centering{=}\ \ \begin{tikzpicture}[cat,scale=0.8]
\node (e) at (0,-0.6) {\small $e$};
\draw[thick] (-0.4,-0.4) -- (0.4,-0.4) -- (0,-0.9) -- cycle;
\draw[thick] (0,-0.4) -- (0,0.5);
\end{tikzpicture}
\quad\quad\quad\quad\quad
\begin{tikzpicture}[cat,scale=0.8]
\node (in) at (0,1.2) {};
\node (out) at (0,-1.8) {};
\ground{gr1}{0.7,0.7}
\ground{gr2}{0.7,-0.3}
\node [black vertex] (gdot1) at (0,0) {};
\node [black vertex] (gdot2) at (0,-1) {};
\draw[thick,rounded corners=2pt] (gdot1) -- (0.7,0.5) -- (gr1);
\draw[thick] (gdot1) -- (in);
\draw[thick,rounded corners=2pt] (gdot2) -- (0.7,-0.5)-- (gr2);
\draw[thick] (gdot2) -- (out);
\draw[thick] (gdot1) -- (gdot2);
\end{tikzpicture}\ \ \ \
\centering{=}\ \ \ \ \begin{tikzpicture}[cat,scale=0.8]
\node (in) at (0,1.2) {};
\node (out) at (0,-0.8) {};
\ground{gr}{0.7,0.7}
\node [black vertex] (gdot) at (0,0) {};
\draw[thick,rounded corners=2pt] (gdot) -- (0.7,0.5) -- (gr);
\draw[thick] (gdot) -- (in);
\draw[thick] (gdot) -- (out);
\end{tikzpicture}
\]
\end{prop}

\begin{proof} 
The first equality follows from Eq.~(\ref{eq:claselts}) and Eq.~(\ref{eq:normalize}), and  
\[
\begin{tikzpicture}[cat,scale=0.5]
\node (in) at (0,1.2) {};
\node (out) at (0,-1.8) {};
\ground{gr1}{0.7,0.7}
\ground{gr2}{0.7,-0.3}
\node [black small vertex] (gdot1) at (0,0) {};
\node [black small vertex] (gdot2) at (0,-1) {};
\draw[thick,rounded corners=2pt] (gdot1) -- (0.7,0.5) -- (gr1);
\draw[thick] (gdot1) -- (in);
\draw[thick,rounded corners=2pt] (gdot2) -- (0.7,-0.5)-- (gr2);
\draw[thick] (gdot2) -- (out);
\draw[thick] (gdot1) -- (gdot2);
\end{tikzpicture}\ \ 
\centering{=}\ \ \begin{tikzpicture}[cat,scale=0.5]
\node (in) at (0,1.2) {};
\node (out) at (0,-0.8) {};
\ground{gr}{0.7,0.7}
\node [black small vertex] (gdot) at (0,0) {};
\draw[thick,rounded corners=2pt] (gdot) -- (0.7,0.5) -- (gr);
\draw[thick] (gdot) -- (in);
\draw[thick] (gdot) -- (out);
\end{tikzpicture}\ \ 
\centering{\iff}\ \ 
\begin{tikzpicture}[cat,scale=0.5]
\node (out) at (0,-1.8) {};
\node (out2) at (-0.5,-1.8) {};
\ground{gr1}{0.7,0.7}
\ground{gr2}{0.7,-0.3}
\node [black small vertex] (gdot1) at (0,0) {};
\node [black small vertex] (gdot2) at (0,-1) {};
\draw[thick,rounded corners=2pt] (gdot1) -- (0.7,0.5) -- (gr1);
\draw[thick,rounded corners=2pt] (gdot1) -- (0,0.8) -- (-0.5,0.8) -- (out2);
\draw[thick,rounded corners=2pt] (gdot2) -- (0.7,-0.5)-- (gr2);
\draw[thick] (gdot2) -- (out);
\draw[thick] (gdot1) -- (gdot2);
\end{tikzpicture}\ \ 
\centering{=}\ \ \begin{tikzpicture}[cat,scale=0.5]
\node (in) at (-0.5,-0.8) {};
\node (out) at (0,-0.8) {};
\ground{gr}{0.7,0.7}
\node [black small vertex] (gdot) at (0,0) {};
\draw[thick,rounded corners=2pt] (gdot) -- (0.7,0.5) -- (gr);
\draw[thick,rounded corners=2pt] (gdot1) -- (0,0.8) -- (-0.5,0.8) -- (in);
\draw[thick] (gdot) -- (out);
\end{tikzpicture}\ \ 
\centering{\ \stackrel{(\ref{eq:environment2})}{\Longleftrightarrow} \ }\ \ 
\begin{tikzpicture}[cat,scale=0.5]
\node (out) at (0,-1.8) {};
\node (out2) at (-0.5,-1.8) {};
\node [black small vertex] (gdot1) at (0,-0.4) {};
\node [black small vertex] (gdot2) at (0,-1) {};
\draw[thick,rounded corners=2pt] (gdot1) -- (0,-0.1) -- (-0.5,-0.1) -- (out2);
\draw[thick] (gdot2) -- (out);
\draw[thick] (gdot1) -- (gdot2);
\node (out4) at (0,1.8) {};
\node (out3) at (-0.5,1.8) {};
\node [black small vertex] (gdot3) at (0,0.4) {};
\node [black small vertex] (gdot4) at (0,1) {};
\draw[thick,rounded corners=2pt] (gdot1) -- (0.3,0) -- (gdot3);
\draw[thick,rounded corners=2pt] (gdot3) -- (0,0.1) -- (-0.5,0.1) -- (out3);
\draw[thick,rounded corners=2pt] (gdot2) -- (0.5,-0.3)--  (0.5,0.3) -- (gdot4);
\draw[thick] (gdot4) -- (out4);
\draw[thick] (gdot3) -- (gdot4);
\end{tikzpicture}\ \ 
\centering{=}\ \ \begin{tikzpicture}[cat,scale=0.5]
\node (out) at (0,-1.2) {};
\node (out2) at (-0.5,-1.2) {};
\node [black small vertex] (gdot1) at (0,-0.4) {};
\draw[thick,rounded corners=2pt] (gdot1) -- (0,-0.1) -- (-0.5,-0.1) -- (out2);
\draw[thick] (gdot1) -- (out);
\node (out4) at (0,1.2) {};
\node (out3) at (-0.5,1.2) {};
\node [black small vertex] (gdot3) at (0,0.4) {};
\draw[thick,rounded corners=2pt] (gdot1) -- (0.3,0) -- (gdot3);
\draw[thick,rounded corners=2pt] (gdot3) -- (0,0.1) -- (-0.5,0.1) -- (out3);
\draw[thick] (gdot3) -- (out4);
\end{tikzpicture}
\]
where the last equality holds  due to the spider normal form theorem.  
\end{proof}

We can now construct a \em measurement \em as follows:
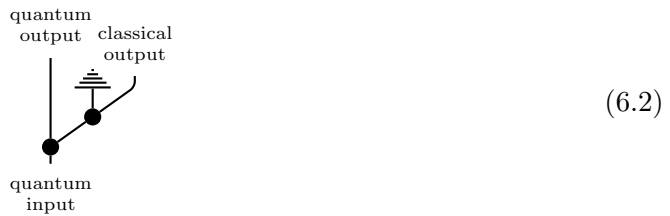
\beq\label{eq:meas}
\begin{tikzpicture}[cat,scale=0.8]
\node (out1) at (0,2) { {\tiny \begin{tabular}{c}quantum\\output\end{tabular}}};
\node (in) at (0,-0.8) { {\tiny \begin{tabular}{c}quantum\\input\end{tabular}}};
\node (out2) at (1.4,1.7) { {\tiny \begin{tabular}{c}classical\\output\end{tabular}}};
\node [black vertex] (gdot2) at (0.7,0.5) {};
\ground{gr}{0.7,01}
\node [black vertex] (gdot1) at (0,0) {};
\draw[thick] (gdot2) -- (gdot1);
\draw[thick] (gdot2) -- (gr);
\draw[thick,rounded corners=2pt] (gdot2) -- (1.4,1) -- (out2);
\draw[thick] (gdot1) -- (out1);
\draw[thick] (gdot1) -- (in);
\end{tikzpicture}
\eeq
i.e.~it copies  the quantum data into a classical channel. A \em destructive measurement \em is obtained by  `feeding the quantum output   itself  into the environment'. Proposition \ref{prop:idempotence} then yields:
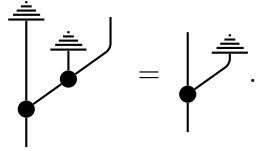
\[
\begin{tikzpicture}[cat,scale=0.8]
\ground{out1}{0,1.5}
\node (in) at (0,-0.8) {};
\node (out2) at (1.4,1.7) {};
\node [black vertex] (gdot2) at (0.7,0.5) {};
\ground{gr}{0.7,01}
\node [black vertex] (gdot1) at (0,0) {};
\draw[thick] (gdot2) -- (gdot1);
\draw[thick] (gdot2) -- (gr);
\draw[thick,rounded corners=2pt] (gdot2) -- (1.4,1) -- (out2);
\draw[thick] (gdot1) -- (out1);
\draw[thick] (gdot1) -- (in);
\end{tikzpicture}
~~~~~=~~~~~
\begin{tikzpicture}[cat,scale=0.8]
\node (in) at (0,1.2) {};
\node (out) at (0,-0.8) {};
\ground{gr}{0.7,0.7}
\node [black vertex] (gdot) at (0,0) {};
\draw[thick,rounded corners=2pt] (gdot) -- (0.7,0.5)-- (gr);
\draw[thick] (gdot) -- (in);
\draw[thick] (gdot) -- (out);
\end{tikzpicture}.
\]
and the resulting shape of the destructive measurement is then:
\beq\label{eq:distmeas}
\begin{tikzpicture}[cat,scale=0.8]
\node (in) at (0,1.2) {  {\tiny \begin{tabular}{c}classical\\output\end{tabular}}};
\node (out) at (0,-0.8) { {\tiny \begin{tabular}{c}quantum\\input\end{tabular}}};
\ground{gr}{0.7,0.7}
\node [black vertex] (gdot) at (0,0) {};
\draw[thick,rounded corners=2pt] (gdot) -- (0.7,0.5)-- (gr);
\draw[thick] (gdot) -- (in);
\draw[thick] (gdot) -- (out);
\end{tikzpicture}. 
\eeq
Note here in particular that  destructive measurements and classical channels are `semantically equivalent'.  Similarly, by the  spider normal form  we have:
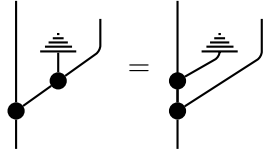
\[
\begin{tikzpicture}[cat,scale=0.8]
\node (out1) at (0,2) {};
\node (in) at (0,-0.8) {};
\node (out2) at (1.4,1.7) {};
\node [black vertex] (gdot2) at (0.7,0.5) {};
\ground{gr}{0.7,01}
\node [black vertex] (gdot1) at (0,0) {};
\draw[thick] (gdot2) -- (gdot1);
\draw[thick] (gdot2) -- (gr);
\draw[thick,rounded corners=2pt] (gdot2) -- (1.4,1) -- (out2);
\draw[thick] (gdot1) -- (out1);
\draw[thick] (gdot1) -- (in);
\end{tikzpicture}
~~~~~=~~~~~
\begin{tikzpicture}[cat,scale=0.8]
\node (out1) at (0,2) {};
\node (in) at (0,-0.8) {};
\node (out2) at (1.4,1.7) {};
\node [black vertex] (gdot2) at (0,0.5) {};
\ground{gr}{0.7,01}
\node [black vertex] (gdot1) at (0,0) {};
\draw[thick] (gdot2) -- (gdot1);
\draw[thick,rounded corners=2pt] (gdot2) -- (0.7,0.9) --(gr);
\draw[thick,rounded corners=2pt] (gdot1) -- (1.4,0.9) -- (out2);
\draw[thick] (gdot1) -- (out1);
\draw[thick] (gdot1) -- (in);
\end{tikzpicture}
\]
so the quantum output of a measurement is `semantically equivalent' to its classical output, which captures change of the quantum state to an eigenstate. More generally,   as a consequence of the structural power  of the spider normal form theorem, classicality `semantically spreads through a diagram'. 

\begin{exa}
We now illustrate the above exposed diagrammatic analysis on the concrete example of measurement of a qubit.  For $|\psi\rangle = \psi_0 |0\rangle + \psi_1 |1\rangle$ to which we apply the morphism of  of Eq.~(\ref{eq:meas}) which we assume to be in the computational basis,  
the first `copying' operation yields $ \psi_0 |00\rangle + \psi_1 |11\rangle$, the second one yields $\psi_0 |000\rangle + \psi_1 |111\rangle$ and the effect of the environment yields, now necessarily in density matrix terms,  $|\psi_0|^2 |00\rangle\langle 00| + |\psi_1|^2 |11\rangle\langle 11|$.
\end{exa}

Note that by idempotence of $C_\Xi$ it also follows that $\bot_A=\top_A^\dagger$ 
is a classical element, and in particular, that this does not depend on the choice of $\Xi$.
We will call $\bot_A$ (unnormalized) \em maximal mixedness\em.  

\begin{exa}
In  $\D$  we indeed  have that $\bot_\mathcal{ H}$ is diagonal in any basis.
\end{exa}

We call a morphism $f:A\to B$ \em disconnected \em if it factors along $\II$, that is,  if $f= \psi\circ \pi$ for some  $\psi:\II\to B$ and $\pi:A\to \II$.  In the graphical representation 
we indeed obtain a disconnected picture in this case:
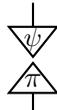
\[
\begin{tikzpicture}[cat,scale=0.8]
\node (e) at (0,-0.6) {\small $\psi$};
\draw[thick] (-0.4,-0.4) -- (0.4,-0.4) -- (0,-0.9) -- cycle;
\draw[thick] (0,-0.4) -- (0,0);
\node (e) at (0,-1.3) {$\pi$};
\draw[thick] (-0.4,-1.5) -- (0.4,-1.5) -- (0,-1) -- cycle;
\draw[thick] (0,-1.5) -- (0,-1.8);
\end{tikzpicture}
\]
The topological disconnectedness physically stands for the fact that there is no  information flowing  from the input to the output.

\begin{rem}
For non-trivial categories the morphisms $\top_A$ cannot be pure; if they would be pure then setting $f:=\top_A$ and $g:=1_A$ in  Eq.~\ref{eq:environment2}, the righthandside becomes $1_{\rm I}\circ \top_A=\top_A\circ 1_A$, which holds, and hence also the lefthandside holds: $\bot_A\circ\top_A=1_A\circ 1_A= 1_A$. That is, the identity is `disconnected'. This is obviously in conflict with the intuition that through a straight wire information flows without being modified, so one expects bad things to happen.  Indeed,  for any $f:A\to B$ we now have 
$f=1_B\circ f\circ 1_A= \bot_B\circ\top_B\circ f\circ \bot_A\circ\top_A=s\cdot \bot_B\circ\top_A$ with $s= \top_B\circ f \bot_A:\II\to\II$ a scalar. That is, any morphism is disconnected!
\end{rem}

If we introduce $H$ between $C_\Xi$ and itself, we obtain `complementary behaviors'.  
The first equality of the following proposition implies that a measurement turns a pure classical element of a complementary measurement in maximal mixedness, i.e.~any outcome is equally probable  for that measurement (cf.~`unbiasedness').  The second one implies that there is no dataflow from the input to the output when we compose complementary measurements.

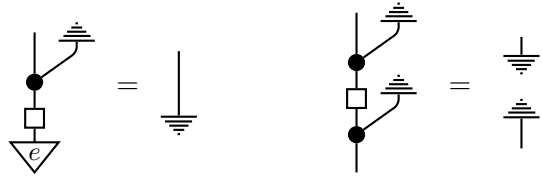
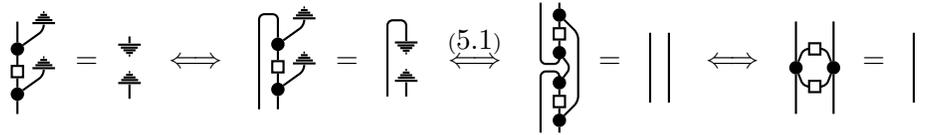
\begin{prop}\label{pr:compunbchannels}
\ \vspace{-5.5mm}\[
\begin{tikzpicture}[cat,scale=0.8]
\node (in) at (0,1) {};
\ground{gr1}{0.7,0.7}
\node[h vertex] (h) at (0,-0.6) {};
\node (e) at (0,-1.2) {\small $e$};
\draw[thick] (0,-1) -- (h);
\draw[thick] (-0.4,-1) -- (0.4,-1) -- (0,-1.5) -- cycle;
\node [black vertex] (gdot1) at (0,0) {};
\draw[thick,rounded corners=2pt] (gdot1) -- (0.7,0.5)-- (gr1);
\draw[thick] (gdot1) -- (in);
\draw[thick] (gdot1) -- (h);
\end{tikzpicture}\ \ 
\centering{=}\ \ \begin{tikzpicture}[cat,scale=0.8]
\maxmixed{gr}{0,-0.6}
\draw[thick] (gr) -- (0,0.5);
\end{tikzpicture}
\quad\quad\quad\quad\quad
\begin{tikzpicture}[cat,scale=0.8]
\node (in) at (0,1.2) {};
\node (out) at (0,-1.8) {};
\ground{gr1}{0.7,0.9}
\node[h vertex] (h) at (0,-0.4) {};
\ground{gr2}{0.7,-0.3}
\node [black vertex] (gdot1) at (0,0.2) {};
\node [black vertex] (gdot2) at (0,-1) {};
\draw[thick,rounded corners=2pt] (gdot1) --  (0.7,0.7) -- (gr1);
\draw[thick] (gdot1) -- (in);
\draw[thick,rounded corners=2pt] (gdot2) -- (0.7,-0.5) -- (gr2);
\draw[thick] (gdot2) -- (out);
\draw[thick] (gdot1) -- (h) -- (gdot2);
\end{tikzpicture}\ \ \ 
\centering{=}\ \ \ \begin{tikzpicture}[cat,scale=0.8]
\node (in) at (0,1) {};
\maxmixed{gr1}{0,0.5}
\ground{gr2}{0,-0.5}
\node (out) at (0,-1.2) {};
\draw[thick] (in) -- (gr1);
\draw[thick] (gr2) -- (out);
\end{tikzpicture}
\]
\end{prop}

\begin{proof} We have:
\[
\begin{tikzpicture}[cat,scale=0.5]
\node (in) at (0,1.2) {};
\node (out) at (0,-1.8) {};
\ground{gr1}{0.7,0.9}
\node[h small vertex] (h) at (0,-0.4) {};
\ground{gr2}{0.7,-0.3}
\node [black small vertex] (gdot1) at (0,0.2) {};
\node [black small vertex] (gdot2) at (0,-1) {};
\draw[thick,rounded corners=2pt] (gdot1) --  (0.7,0.7) -- (gr1);
\draw[thick] (gdot1) -- (in);
\draw[thick,rounded corners=2pt] (gdot2) -- (0.7,-0.5) -- (gr2);
\draw[thick] (gdot2) -- (out);
\draw[thick] (gdot1) -- (h) -- (gdot2);
\end{tikzpicture}\ \ 
\centering{=}\ \ \begin{tikzpicture}[cat,scale=0.5]
\node (in) at (0,1) {};
\maxmixed{gr1}{0,0.5}
\ground{gr2}{0,-0.5}
\node (out) at (0,-1.2) {};
\draw[thick] (in) -- (gr1);
\draw[thick] (gr2) -- (out);
\end{tikzpicture}\ \ 
\centering{\iff}\ \ %
\begin{tikzpicture}[cat,scale=0.5]
\node (out) at (0,-1.8) {};
\node (in) at (-0.5,-1.8) {};
\ground{gr1}{0.7,0.9}
\node[h small vertex] (h) at (0,-0.4) {};
\ground{gr2}{0.7,-0.3}
\node [black small vertex] (gdot1) at (0,0.2) {};
\node [black small vertex] (gdot2) at (0,-1) {};
\draw[thick,rounded corners=2pt] (gdot1) --  (0.7,0.7) -- (gr1);
\draw[thick,rounded corners=2pt] (gdot1) -- (0,1)  -- (-0.5, 1) -- (in);
\draw[thick,rounded corners=2pt] (gdot2) -- (0.7,-0.5) -- (gr2);
\draw[thick] (gdot2) -- (out);
\draw[thick] (gdot1) -- (h) -- (gdot2);
\end{tikzpicture}\ \ 
\centering{=}\ \ \begin{tikzpicture}[cat,scale=0.5]
\node (in) at (-0.5,-1.2) {};
\maxmixed{gr1}{0,0.5}
\ground{gr2}{0,-0.5}
\node (out) at (0,-1.2) {};
\draw[thick,rounded corners=2pt]  (in) --(-0.5,1.1) --(0,1.1)-- (gr1);
\draw[thick] (gr2) -- (out);
\end{tikzpicture}\ \ 
\centering{\  \stackrel{(\ref{eq:environment2})}{\Longleftrightarrow} \ }\ \ 
\begin{tikzpicture}[cat,scale=0.5]
\node (out) at (0,-2) {};
\node (out2) at (-0.5,-2) {};
\node [black small vertex] (gdot1) at (0,-0.4) {};
\node [h small vertex] (h1) at (0,-0.9) {};
\node [black small vertex] (gdot2) at (0,-1.4) {};
\draw[thick,rounded corners=2pt] (gdot1) -- (0,-0.1) -- (-0.5,-0.1) -- (out2);
\draw[thick] (gdot2) -- (out);
\draw[thick] (gdot1) -- (h1) -- (gdot2);
\node (out4) at (0,2) {};
\node (out3) at (-0.5,2) {};
\node [black small vertex] (gdot3) at (0,0.4) {};
\node [h small vertex] (h2) at (0,0.9) {};
\node [black small vertex] (gdot4) at (0,1.4) {};
\draw[thick,rounded corners=2pt] (gdot1) -- (0.3,0) -- (gdot3);
\draw[thick,rounded corners=2pt] (gdot3) -- (0,0.1) -- (-0.5,0.1) -- (out3);
\draw[thick,rounded corners=2pt] (gdot2) -- (0.5,-0.9)--  (0.5,0.9) -- (gdot4);
\draw[thick] (gdot4) -- (out4);
\draw[thick] (gdot3) -- (h2) -- (gdot4);
\end{tikzpicture}\ \ 
\centering{=}\ \ \begin{tikzpicture}[cat,scale=0.5]
\node (out) at (0,-1.2) {};
\node (out2) at (-0.5,-1.2) {};
\node (out4) at (0,1.2) {};
\node (out3) at (-0.5,1.2) {};
\draw[thick] (out) -- (out4);
\draw[thick] (out2) -- (out3);
\end{tikzpicture}
\ \ 
\centering{\iff}\ \ 
\begin{tikzpicture}[cat,scale=0.5]
\node (out) at (0,-1.5) {};
\node (out2) at (-1,-1.5) {};
\node (out4) at (0,1.5) {};
\node (out3) at (-1,1.5) {};
\node [black small vertex] (gdot1) at (-1,0) {};
\node [black small vertex] (gdot2) at (0,0) {};
\node [h small vertex] (h1) at (-0.5,-0.5) {};
\node [h small vertex] (h2) at (-0.5,0.5) {};
\draw[thick] (out) -- (gdot2) --  (out4);
\draw[thick] (out2) -- (gdot1) -- (out3);
\draw[thick,rounded corners=2pt] (gdot1) -- (-0.8,-0.5) -- (h1) -- (-0.2,-0.5) -- (gdot2) ;
\draw[thick,rounded corners=2pt] (gdot1) -- (-0.8,0.5) -- (h2) -- (-0.2,0.5) -- (gdot2) ;
\end{tikzpicture}
\ \ 
\centering{=}\ \ 
\begin{tikzpicture}[cat,scale=0.5]
\node (out) at (0,-1.2) {};
\node (out2) at (-0.5,-1.2) {};
\node (out4) at (0,1.2) {};
\node (out3) at (-0.5,1.2) {};
\draw[thick] (out) -- (out4);
\draw[thick] (out2) -- (out3);
\end{tikzpicture}
\]
The first equation is derived from the second one and Proposition \ref{prop:idempotence}:
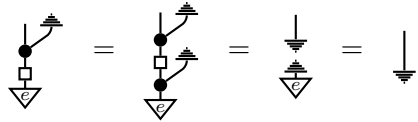
\[
\begin{tikzpicture}[cat,scale=0.5]
\node (in) at (0,1) {};
\ground{gr1}{0.7,0.7}
\node[h small vertex] (h) at (0,-0.6) {};
\node (e) at (0,-1.2) {\tiny $e$};
\draw[thick] (0,-1) -- (h);
\draw[thick] (-0.4,-1) -- (0.4,-1) -- (0,-1.5) -- cycle;
\node [black small vertex] (gdot1) at (0,0) {};
\draw[thick,rounded corners=2pt] (gdot1) -- (0.7,0.5)-- (gr1);
\draw[thick] (gdot1) -- (in);
\draw[thick] (gdot1) -- (h);
\end{tikzpicture}
\ \ \ 
\centering{=}\ \ \ 
\begin{tikzpicture}[cat,scale=0.5]
\node (in) at (0,1.2) {};
\ground{gr1}{0.7,0.9}
\node[h small vertex] (h) at (0,-0.4) {};
\ground{gr2}{0.7,-0.3}
\node [black small vertex] (gdot1) at (0,0.2) {};
\node [black small vertex] (gdot2) at (0,-1) {};
\draw[thick,rounded corners=2pt] (gdot1) --  (0.7,0.7) -- (gr1);
\draw[thick] (gdot1) -- (in);
\draw[thick,rounded corners=2pt] (gdot2) -- (0.7,-0.5) -- (gr2);
\draw[thick] (gdot1) -- (h) -- (gdot2);
\node (e) at (0,-1.6) {\tiny $e$};
\draw[thick] (0,-1.4) -- (gdot2);
\draw[thick] (-0.4,-1.4) -- (0.4,-1.4) -- (0,-1.9) -- cycle;
\end{tikzpicture}
\ \ \ 
\centering{=}\ \ \ \begin{tikzpicture}[cat,scale=0.5]
\maxmixed{mx}{0,-0.4}
\draw[thick] (mx) -- (0,0.3);
\ground{gr2}{0,-1.2}
\node (e) at (0,-1.6) {\tiny $e$};
\draw[thick] (0,-1.4) -- (gr2);
\draw[thick] (-0.4,-1.4) -- (0.4,-1.4) -- (0,-1.9) -- cycle;
\end{tikzpicture}
\ \ \ 
\centering{=}\ \ \ \begin{tikzpicture}[cat,scale=0.5]
\maxmixed{gr}{0,-0.6}
\draw[thick] (gr) -- (0,0.5);
\end{tikzpicture}
\]
\end{proof}

\subsection{General classical control operations}\label{subsec:cont}

In diagrammatic terms, a morphism $U: A\to A$ is unitary if:
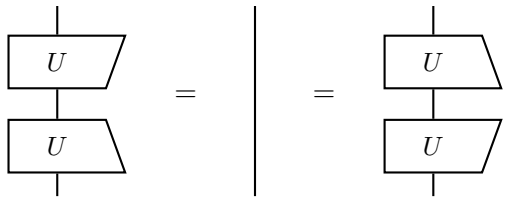
\[\begin{tikzpicture}[cat,scale=0.8]
\node (out) at (0,3.3) {};
\node[dagger morph] (ud) at (0,0.8) {\small $~U~$};
\node[morph] (u) at (0,2.2) {\small $~U~$};
\node (in) at (0,-0.2) {};
\draw[thick] (in) -- (ud) --(u) -- (out);
\end{tikzpicture}\ \ \ \ \
\centering{=}\ \ \ \ \ 
\begin{tikzpicture}[cat,scale=0.8]
\node (out) at (0,3.3) {};
\node (in) at (0,-0.2) {};
\draw[thick] (in) -- (out);
\end{tikzpicture}\ \ \ \ \ 
\centering{=}\ \ \ \ \ 
\begin{tikzpicture}[cat,scale=0.8]
\node (out) at (0,3.3) {};
\node[morph] (u) at (0,0.8) {\small $~U~$};
\node[dagger morph] (ud) at (0,2.2) {\small $~U~$};
\node (in) at (0,-0.2) {};
\draw[thick] (in) -- (u) --(ud) -- (out);
\end{tikzpicture}
\]
We now define what it means to have a family of unitaries of the same type, `indexed' by a classical structure, that is, a controlled unitary.

\begin{defi}
By a \em controlled unitary \em we mean an operation of the form:
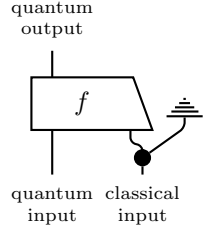
\[\begin{tikzpicture}[cat,scale=0.8]
\node (out) at (-1.5,2.3) { {\tiny \begin{tabular}{c}quantum\\output\end{tabular}}};
\node[dagger morph] (f) at (-1,0.9) {\small $\ \ f\ \ $};
\node (inc) at (0,-0.8) { {\tiny \begin{tabular}{c}classical\\input\end{tabular}}};
\node (inq) at (-1.5,-0.8) { {\tiny \begin{tabular}{c}quantum\\input\end{tabular}}};
\ground{gr}{0.7,0.7}
\node [black vertex] (gdot) at (0,0) {};
\draw[thick,rounded corners=2pt] (gdot) -- (0.7,0.5)-- (gr);
\draw[thick,rounded corners=2pt] (gdot) -- (0,0.2) -- (-0.2,0.25) -- (-0.2,0.46);
\draw[thick] (inq) -- (-1.5,0.46);
\draw[thick] (out) -- (-1.5,1.33);
\draw[thick] (gdot) -- (inc);
\end{tikzpicture}
\]
which `for all classical input values is unitary', that is: 
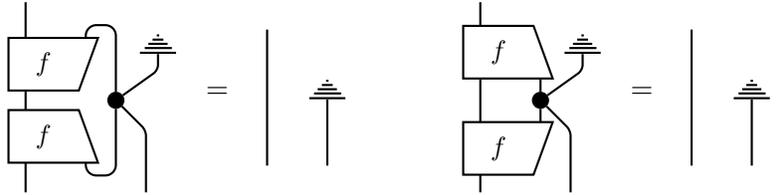
\[
\begin{tikzpicture}[cat,scale=0.8]
\node (out) at (-1.5,3.3) {};
\node[dagger morph] (f) at (-1.2,0.9) {\small $\ f\ $};
\node[morph] (f) at (-1.2,2.1) {\small $\ f\ $};
\node (inc) at (0.5,-0.2) {};
\node (inq) at (-1.5,-0.2) {};
\ground{gr}{0.7,2.3}
\node [black vertex] (gdot) at (0,1.5) {};
\draw[thick,rounded corners=2pt] (gdot) -- (0.7,2)-- (gr);
\draw[thick,rounded corners=4pt] (gdot) -- (0,0.25)-- (-0.5,0.25) -- (-0.5,0.46);
\draw[thick,rounded corners=4pt] (gdot) -- (0,2.75)-- (-0.5,2.75) -- (-0.5,2.55);
\draw[thick] (inq) -- (-1.5,0.46);
\draw[thick] (-1.5,1.65) -- (-1.5,1.33);
\draw[thick] (-1.5,2.55) -- (out);
\draw[thick,rounded corners=2pt] (gdot) -- (0.5,1) -- (inc);
\end{tikzpicture}\ \ \ 
\centering{=}\ \ \ 
\begin{tikzpicture}[cat,scale=0.8]
\ground{gr}{0,0};
\node (inc) at (0,-1.3) {};
\draw[thick] (gr) -- (inc);
\node (inq) at (-1,-1.3) {};
\node (outq) at (-1,1.3) {};
\draw[thick] (inq) -- (outq);
\end{tikzpicture}\qquad\qquad
\begin{tikzpicture}[cat,scale=0.8]
\node (out) at (-1.5,3.3) {};
\node[morph] (f) at (-1.2,0.7) {\small $\ f\ $};
\node[dagger morph] (f) at (-1.2,2.3) {\small $\ f\ $};
\node (inc) at (0,-0.2) {};
\node (inq) at (-1.5,-0.2) {};
\ground{gr}{0.2,2.3}
\node [black vertex] (gdot) at (-0.5,1.5) {};
\draw[thick,rounded corners=2pt] (gdot) -- (0.2,2)-- (gr);
\draw[thick,rounded corners=4pt] (-0.5,1.13) -- (gdot) -- (-0.5,1.85);
\draw[thick] (inq) -- (-1.5,0.26);
\draw[thick] (-1.5,1.85) -- (-1.5,1.13);
\draw[thick] (-1.5,2.75) -- (out);
\draw[thick,rounded corners=2pt] (gdot) -- (0,1) -- (inc);
\end{tikzpicture}\ \ \ 
\centering{=}\ \ \ 
\begin{tikzpicture}[cat,scale=0.8]
\ground{gr}{0,0};
\node (inc) at (0,-1.3) {};
\draw[thick] (gr) -- (inc);
\node (inq) at (-1,-1.3) {};
\node (outq) at (-1,1.3) {};
\draw[thick] (inq) -- (outq);
\end{tikzpicture}\ . 
\]
\end{defi}

\begin{lem}\label{lem:controlex}
The following morphisms  are controlled unitaries: 
\beq\label{eq:contunitary}
\begin{tikzpicture}[cat,scale=0.8]
\node (out) at (-1,1.4) {};
\node[black vertex] (f) at (-1,0.5) {};
\node (inc) at (0,-0.8) {};
\node (inq) at (-1,-0.8) {};
\ground{gr}{0.4,0.7}
\node[black vertex] (gdot) at (0,0) {};
\node[h vertex] (h) at (-0.5,0.25) {};
\draw[thick,rounded corners=2pt] (gdot) -- (0.4,0.5)-- (gr); 
\draw[thick,rounded corners=5pt] (gdot)  -- (h) -- (f);
\draw[thick] (inq) -- (f);
\draw[thick] (out) -- (f);
\draw[thick] (gdot) -- (inc);
\end{tikzpicture}
\qquad\qquad\qquad
\begin{tikzpicture}[cat,scale=0.8]
\node (out) at (-1,1.6) {};
\node[black vertex] (f) at (-1,0.5) {};
\node (inc) at (0,-0.8) {};
\node (inq) at (-1,-0.8) {};
\ground{gr}{0.4,0.7}
\node[black vertex] (gdot) at (0,0) {};
\node[h vertex] (h) at (-0.5,0.25) {};
\node[h vertex] (h1) at (-1,0) {};
\node[h vertex] (h2) at (-1,1) {};
\draw[thick,rounded corners=2pt] (gdot) -- (0.4,0.5)-- (gr); 
\draw[thick,rounded corners=5pt] (gdot)  -- (h) -- (f);
\draw[thick] (inq) -- (h1) -- (f);
\draw[thick] (out) --  (h2) -- (f);
\draw[thick] (gdot) -- (inc);
\end{tikzpicture}
\eeq
\end{lem}

\begin{proof}
We have:
\[
\begin{tikzpicture}[cat,scale=0.5]
\node (out) at (-1.4,1.4) {};
\node[black small vertex] (f) at (-1.4,0.5) {};
\node[black small vertex] (g) at (-1.4,-0.5) {};
\node (inc) at (0,-1.4) {};
\node (inq) at (-1.4,-1.4) {};
\ground{gr}{0.4,0.7}
\node[black small vertex] (gdot) at (0,0) {};
\node[h small vertex] (h) at (-0.7,0.25) {};
\node[h small vertex] (h2) at (-0.7,-0.25) {};
\draw[thick,rounded corners=2pt] (gdot) -- (0.4,0.5)-- (gr);
\draw[thick,rounded corners=5pt] (gdot)  -- (h) -- (f);
\draw[thick,rounded corners=5pt] (gdot)  -- (h2) -- (g);
\draw[thick] (inq) -- (g) -- (f) -- (out);
\draw[thick] (gdot) -- (inc);
\end{tikzpicture}\ \ \ 
\centering{=}\ \ \ 
\begin{tikzpicture}[cat,scale=0.5]
\node (out) at (-1.4,1.4) {};
\node[black small vertex] (f) at (-1.4,0) {};
\node (inc) at (0,-1.4) {};
\node (inq) at (-1.4,-1.4) {};
\ground{gr}{0.4,0.7}
\node[black small vertex] (gdot) at (0,0) {};
\node[h small vertex] (h) at (-0.7,0.3) {};
\node[h small vertex] (h2) at (-0.7,-0.3) {};
\draw[thick,rounded corners=2pt] (gdot) -- (0.4,0.5)-- (gr);
\draw[thick,rounded corners=2pt] (gdot)  -- (-0.4,0.3) -- (h) -- (-1,0.3)-- (f);
\draw[thick,rounded corners=2pt] (gdot)  -- (-0.4,-0.3) -- (h2) -- (-1,-0.3)-- (f);
\draw[thick] (inq)  -- (f) -- (out);
\draw[thick] (gdot) -- (inc);
\end{tikzpicture}\ \ \ 
\centering{=}\ \ \ 
\begin{tikzpicture}[cat,scale=0.5]
\ground{gr}{0,0};
\node (inc) at (0,-1.3) {};
\draw[thick] (gr) -- (inc);
\node (inq) at (-1,-1.3) {};
\node (outq) at (-1,1.3) {};
\draw[thick] (inq) -- (outq);
\end{tikzpicture}
\]
and the remainder of the proof proceeds almost identical.
\end{proof}

\subsection{General non-degenerate measurements}\label{subsec:meas}

We have identified an example of a \em non-degenerate measurement\em, namely the one of the shape (\ref{eq:meas}), and an example of a \em non-degenerate destructive measurement\em, namely the one of the shape (\ref{eq:distmeas}).  Relative to a given classical structure we can define more general non-destructive measurements.  

The following Lemma shows how classical data can be composed in terms of classical structures, where  we conceive classical structures as being specified by a multiplication and its unit, i.e.~$\Xi:=(\Xi_2^1, \Xi_0^1)$ --cf.~Remark \ref{rem:specifyspider}. 

\begin{lem}
If $(\Xi_2^1, \Xi_0^1)$ and $(\tilde{\Xi}_2^1, \tilde{\Xi}_0^1)$ are classical structures on $A$ and $\tilde{A}$ respectively, then the morphisms
\[ 
(\Xi_2^1\otimes \tilde{\Xi}_2^1)\circ (1_A\otimes \sigma_{A, \tilde{A}}\otimes 1_{\tilde{A}}): (A\otimes \tilde{A})\otimes (A\otimes \tilde{A})\to A\otimes \tilde{A}
\]
\[
\Xi_0^1\otimes\tilde{\Xi}_0^1 : {\rm I}\to A\otimes \tilde{A}
\]
define a classical structure on $A\otimes \tilde{A}$, diagrammatically, 
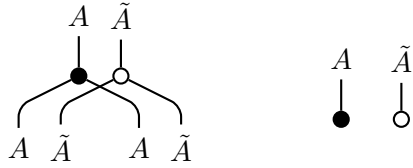
\[
\begin{tikzpicture}[cat,scale=0.8]
\node (in1) at (0,1.5) {$A$};
\node (out1a) at (-1,-0.7) {$A$};
\node (out1b) at (1,-0.7) {$A$};
\node [black vertex] (gdot) at (0,0.5) {};
\draw[thick,rounded corners=4pt] (gdot) -- (1,0) -- (out1b);
\draw[thick,rounded corners=4pt] (gdot) -- (-1,0) -- (out1a);
\draw[thick] (gdot) -- (in1);
\node (in2) at (0.7,1.5) {$\tilde A$};
\node (out2a) at (-0.3,-0.7) {$\tilde A$};
\node (out2b) at (1.7,-0.7) {$\tilde A$};
\node [vertex] (gdot2) at (0.7,0.5) {};
\draw[thick,rounded corners=4pt] (gdot2) -- (1.7,0) -- (out2b);
\draw[thick,rounded corners=4pt] (gdot2) -- (-0.3,0) -- (out2a);
\draw[thick] (gdot2) -- (in2);
\end{tikzpicture}\qquad\qquad
\begin{tikzpicture}[cat,scale=0.8]
\node (in1) at (0,1.5) {$A$};
\node [black vertex] (gdot) at (0,0.5) {};
\draw[thick] (gdot) -- (in1);
\node (in2) at (1,1.5) {$\tilde A$};
\node [vertex] (gdot2) at (1,0.5) {};
\draw[thick] (gdot2) -- (in2);
\end{tikzpicture}
\]
\end{lem}

The canonically corresponding  non-degenerate destructive measurement which extracts this compound classical data from a pair of quantum systems is:
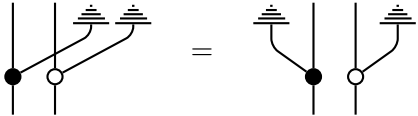
\[
\begin{tikzpicture}[cat,scale=0.8]
\node (in1) at (0,1.5) {};
\node (out1a) at (0,-0.7) {};
\ground{gr1}{1.3,1};
\node [black vertex] (gdot) at (0,0.1) {};
\draw[thick,rounded corners=3pt] (gdot) -- (1.3,0.8) -- (gr1);
\draw[thick,rounded corners=3pt] (gdot) -- (out1a);
\draw[thick] (gdot) -- (in1);
\node (in2) at (0.7,1.5) {};
\node (out2a) at (0.7,-0.7) {};
\ground{gr2}{2,1};
\node [vertex] (gdot2) at (0.7,0.1) {};
\draw[thick,rounded corners=3pt] (gdot2) -- (2,0.8) -- (gr2);
\draw[thick,rounded corners=3pt] (gdot2) -- (out2a);
\draw[thick] (gdot2) -- (in2);
\end{tikzpicture}\ \ \ \
\centering{=}\ \ \ \
\begin{tikzpicture}[cat,scale=0.8]
\node (in1) at (0,1.5) {};
\node (out1a) at (0,-0.7) {};
\ground{gr1}{-0.7,1};
\node [black vertex] (gdot) at (0,0.1) {};
\draw[thick,rounded corners=3pt] (gdot) -- (-0.7,0.6) -- (gr1);
\draw[thick,rounded corners=3pt] (gdot) -- (out1a);
\draw[thick] (gdot) -- (in1);
\node (in2) at (0.7,1.5) {};
\node (out2a) at (0.7,-0.7) {};
\ground{gr2}{1.4,1};
\node [vertex] (gdot2) at (0.7,0.1) {};
\draw[thick,rounded corners=3pt] (gdot2) -- (1.4,0.6) -- (gr2);
\draw[thick,rounded corners=3pt] (gdot2) -- (out2a);
\draw[thick] (gdot2) -- (in2);
\end{tikzpicture}
\]
When transforming the quantum data by means of a unitary we obtain the general form of a non-degenerate distructive measurement on a pair of systems:   
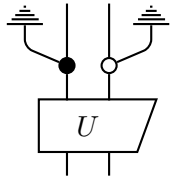
\[
\begin{tikzpicture}[cat,scale=0.8]
\node (in1) at (0,1.5) {};
\node (out1a) at (0,-1.7) {};
\ground{gr1}{-0.7,1};
\node [black vertex] (gdot) at (0,0.3) {};
\draw[thick,rounded corners=3pt] (gdot) -- (-0.7,0.6) -- (gr1);
\draw[thick,rounded corners=3pt] (gdot) -- (0,-0.28) ;
\draw[thick,rounded corners=3pt]  (0,-1.15) -- (out1a);
\draw[thick] (gdot) -- (in1);
\node (in2) at (0.7,1.5) {};
\node (out2a) at (0.7,-1.7) {};
\ground{gr2}{1.4,1};
\node [vertex] (gdot2) at (0.7,0.3) {};
\draw[thick,rounded corners=3pt] (gdot2) -- (1.4,0.6) -- (gr2);
\draw[thick,rounded corners=3pt] (gdot2) -- (0.7,-0.28) ;
\draw[thick,rounded corners=3pt]  (0.7,-1.15) -- (out2a);
\draw[thick] (gdot2) -- (in2);
\node [morph] (g) at (0.35,-0.7) {$~U~$};
\end{tikzpicture}
\]
    
\begin{defi}
Given a classical structure on $A$, a \em non-degenerate measurement \em on $n$ systems of type $A$ is a morphism of the form:
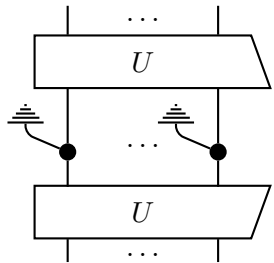
\[
\begin{tikzpicture}[cat,scale=0.8]
\node (in1) at (0,2.9) {};
\node (out1a) at (0,-1.7) {};
\ground{gr1}{-0.7,0.8};
\node (inn) at (2.5,2.9) {};
\node (outna) at (2.5,-1.7) {};
\ground{grn}{1.8,0.8};
\node [black vertex] (gdot) at (0,0.3) {};
\node [black vertex] (gdotn) at (2.5,0.3) {};
\draw[thick,rounded corners=3pt] (gdot) -- (-0.7,0.6) -- (gr1);
\draw[thick,rounded corners=3pt] (gdot) -- (0,-0.28) ;
\draw[thick,rounded corners=3pt]  (0,-1.15) -- (out1a);
\draw[thick,rounded corners=3pt] (gdotn) -- (1.8,0.6) -- (grn);
\draw[thick,rounded corners=3pt] (gdotn) -- (2.5,-0.28) ;
\draw[thick,rounded corners=3pt]  (2.5,-1.15) -- (outna);
\draw[thick] (gdot) -- (0,1.35);
\draw[thick] (0,2.25) -- (in1);
\draw[thick] (gdotn) -- (2.5,1.35);
\draw[thick] (2.5,2.25) -- (inn);
\node [morph] (g) at (1.25,-0.7) {$\ \ \ \ U\ \ \ \ $};
\node [dagger morph] (gd) at (1.25,1.8) {$\ \ \ \ U\ \ \ \ $};
\node (xxx) at (1.25,-1.4) {$\ldots$};
\node (yyy) at (1.25,0.4) {$\ldots$};
\node (yyy) at (1.25,2.5) {$\ldots$};
\end{tikzpicture}
\]
where $U:A\otimes \ldots \otimes  A\to A\otimes \ldots \otimes  A$ is an arbitrary unitary, and 
a corresponding \em non-degenerate destructive measurement \em is a morphism  of the form:
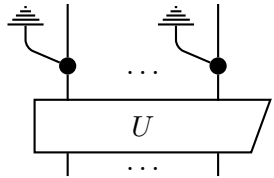
\[
\begin{tikzpicture}[cat,scale=0.8]
\node (in1) at (0,1.5) {};
\node (out1a) at (0,-1.7) {};
\ground{gr1}{-0.7,1};
\node (inn) at (2.5,1.5) {};
\node (outna) at (2.5,-1.7) {};
\ground{grn}{1.8,1};
\node [black vertex] (gdot) at (0,0.3) {};
\node [black vertex] (gdotn) at (2.5,0.3) {};
\draw[thick,rounded corners=3pt] (gdot) -- (-0.7,0.6) -- (gr1);
\draw[thick,rounded corners=3pt] (gdot) -- (0,-0.28) ;
\draw[thick,rounded corners=3pt]  (0,-1.15) -- (out1a);
\draw[thick,rounded corners=3pt] (gdotn) -- (1.8,0.6) -- (grn);
\draw[thick,rounded corners=3pt] (gdotn) -- (2.5,-0.28) ;
\draw[thick,rounded corners=3pt]  (2.5,-1.15) -- (outna);
\draw[thick] (gdot) -- (in1);
\draw[thick] (gdotn) -- (inn);
\node [morph] (g) at (1.25,-0.7) {$\ \ \ \ U\ \ \ \ $};
\node (xxx) at (1.25,-1.4) {$\ldots$};
\node (yyy) at (1.25,0.2) {$\ldots$};
\end{tikzpicture}
\]
\end{defi}

\begin{rem}
One can also define more general kinds of measurements, namely degenerate ones and non-projective ones, which involves defining projective measurements without reference to unitaries.  This can be done in straightforward analogy to how this was done in  \cite{CPav} and \cite{CPaqPav}.  
\end{rem}

\begin{lem}\label{lem:CNOTunitary}
The following morphism is unitary:
\beq
CNOT \ \ \ := \ \ \ \begin{tikzpicture}[cat,scale=0.8]
\node (out) at (-1.4,1.4) {};
\node (out2) at (0,1.4) {};
\node[black vertex] (f) at (-1.4,0) {};
\node (inc) at (0,-1.4) {};
\node (inq) at (-1.4,-1.4) {};
\node[black vertex] (gdot) at (0,0) {};
\node[h vertex] (h) at (-0.7,0) {};
\node[h vertex] (h1) at (0,-0.7) {};
\node[h vertex] (h2) at (0,0.7) {};
\draw[thick,rounded corners=2pt] (gdot)  -- (h) -- (f);
\draw[thick] (inq)  -- (f) -- (out);
\draw[thick] (out2)-- (h2) -- (gdot) --(h1)-- (inc);
\end{tikzpicture}
\eeq
\end{lem}

\begin{proof}
\[
\begin{tikzpicture}[cat,scale=0.5]
\node (out) at (-1.4,2) {};
\node (gr) at (0,2) {};
\node[black small vertex] (f) at (-1.4,0.7) {};
\node[black small vertex] (g) at (-1.4,-0.7) {};
\node (inc) at (0,-2) {};
\node (inq) at (-1.4,-2) {};
\node[black small vertex] (gdot) at (0,0.7) {};
\node[black small vertex] (gdot2) at (0,-0.7) {};
\node[h small vertex] (h) at (-0.7,0.7) {};
\node[h small vertex] (h2) at (-0.7,-0.7) {};
\node[h small vertex] (h3) at (0,-0.25) {};
\node[h small vertex] (h6) at (0,0.25) {};
\node[h small vertex] (h4) at (0,-1.3) {};
\node[h small vertex] (h5) at (0,1.3) {};
\draw[thick,rounded corners=2pt] (gdot) -- (h5) -- (gr);
\draw[thick,rounded corners=5pt] (gdot)  -- (h) -- (f);
\draw[thick,rounded corners=5pt] (gdot2)  -- (h2) -- (g);
\draw[thick] (inq) -- (g) -- (f) -- (out);
\draw[thick] (gdot2) -- (h4) -- (inc);
\draw[thick] (gdot2) -- (h3) -- (h6) -- (gdot);
\end{tikzpicture}\ \ \ 
\centering{=}\ \ \ \begin{tikzpicture}[cat,scale=0.5]
\node (out) at (-1.4,2) {};
\node (gr) at (0,2) {};
\node[black small vertex] (f) at (-1.4,0.7) {};
\node[black small vertex] (g) at (-1.4,-0.7) {};
\node (inc) at (0,-2) {};
\node (inq) at (-1.4,-2) {};
\node[black small vertex] (gdot) at (0,0.7) {};
\node[black small vertex] (gdot2) at (0,-0.7) {};
\node[h small vertex] (h) at (-0.7,0.7) {};
\node[h small vertex] (h2) at (-0.7,-0.7) {};
\node[h small vertex] (h4) at (0,-1.3) {};
\node[h small vertex] (h5) at (0,1.3) {};
\draw[thick,rounded corners=2pt] (gdot) -- (h5) -- (gr);
\draw[thick,rounded corners=5pt] (gdot)  -- (h) -- (f);
\draw[thick,rounded corners=5pt] (gdot2)  -- (h2) -- (g);
\draw[thick] (inq) -- (g) -- (f) -- (out);
\draw[thick] (gdot) -- (h4) -- (inc);
\end{tikzpicture}\ \ \ 
\centering{=}\ \ \ 
\begin{tikzpicture}[cat,scale=0.5]
\node (out) at (-1.4,1.4) {};
\node[h small vertex] (h4) at (0,-0.6) {};
\node[h small vertex] (h5) at (0,0.6) {};
\node (gr) at (0,1.4) {};
\node[black small vertex] (f) at (-1.4,0) {};
\node (inc) at (0,-1.4) {};
\node (inq) at (-1.4,-1.4) {};
\node[black small vertex] (gdot) at (0,0) {};
\node[h small vertex] (h) at (-0.7,0.3) {};
\node[h small vertex] (h2) at (-0.7,-0.3) {};
\draw[thick,rounded corners=2pt] (gdot) -- (h5) -- (gr);
\draw[thick,rounded corners=2pt] (gdot)  -- (-0.4,0.3) -- (h) -- (-1,0.3)-- (f);
\draw[thick,rounded corners=2pt] (gdot)  -- (-0.4,-0.3) -- (h2) -- (-1,-0.3)-- (f);
\draw[thick] (inq)  -- (f) -- (out);
\draw[thick] (gdot) -- (h4) --  (inc);
\end{tikzpicture}\ \ \ 
\centering{=}\ \ \ 
\begin{tikzpicture}[cat,scale=0.5]
\node (out) at (-1,1.4) {};
\node[h small vertex] (h4) at (0,-0.4) {};
\node[h small vertex] (h5) at (0,0.4) {};
\node (gr) at (0,1.4) {};
\node (inc) at (0,-1.4) {};
\node (inq) at (-1,-1.4) {};
\draw[thick] (inq)  -- (out);
\draw[thick] (gr) -- (h5) -- (h4) --  (inc);
\end{tikzpicture}\ \ \ 
\centering{=}\ \ \ 
\begin{tikzpicture}[cat,scale=0.5]
\node (out) at (-1,1.4) {};
\node (gr) at (0,1.4) {};
\node (inc) at (0,-1.4) {};
\node (inq) at (-1,-1.4) {};
\draw[thick] (inq)  -- (out);
\draw[thick] (gr) --  (inc);
\end{tikzpicture}
\]
\end{proof} 

\begin{cor}
The following  morphism is a non-degenerate   destructive measurement on a pair of systems of the same type $A$:
\[
Bell\mbox{-}Meas \ \ \ := \ \ \ \begin{tikzpicture}[cat,scale=0.8]
\node (out) at (-1.4,2.4) {};
\node (out2) at (0,2.4) {};
\node[black vertex] (f) at (-1.4,0) {};
\node[black vertex] (g1) at (-1.4,1.4) {};
\node[black vertex] (g2) at (0,1.4) {};
\ground{gr1}{-2,2};
\ground{gr2}{0.6,2};
\node (inc) at (0,-1.4) {};
\node (inq) at (-1.4,-1.4) {};
\node[black vertex] (gdot) at (0,0) {};
\node[h vertex] (h) at (-0.7,0) {};
\node[h vertex] (h1) at (0,-0.7) {};
\node[h vertex] (h2) at (0,0.7) {};
\node[h vertex] (h3) at (-1.4,0.7) {};
\draw[thick,rounded corners=2pt] (gdot)  -- (h) -- (f);
\draw[thick] (inq)  -- (f) -- (h3) -- (out);
\draw[thick] (out2)-- (h2) -- (gdot) --(h1)-- (inc);
\draw[thick,rounded corners=2pt] (gr1)  -- (-2,1.8) -- (g1);
\draw[thick,rounded corners=2pt] (gr2)  -- (0.6,1.8) -- (g2);
\end{tikzpicture}
\]
\end{cor}

\subsection{Interpretation of graphical elements in $\D$}\label{sec:Densinterp}

The following tables translate the graphical language to Hilbert space quantum theory for the specific case of qubits.    It is this translation which connects that diagrammatic presentation of the protocols in the following section to how one finds them usually described in textbooks. \medskip

\begin{center}
{\bf (pure)  states \& effects:}\\
\begin{tabular}{c|c|c|c|c|}
\hline
Notation:  & \state & \effect & \statebis &\effectbis \\
\hline
$\Dpure$:   & $1\mapsto 2\cdot|+\rangle\langle+|$& $\rho \mapsto 2\cdot\langle+|\rho |+\rangle$& $1\mapsto 2\cdot|0\rangle\langle0|$& $\rho \mapsto 2\cdot\langle0|\rho |0\rangle$\\
\hline
\end{tabular} 
\end{center}

\begin{center}
\begin{tabular}{c|c|c|}
\hline
Notation:  &  \cup & \cap \\ 
\hline
$\Dpure$:   &  $1\mapsto (|00\rangle  + |11\rangle)(\langle00| +\langle11|)  $ & $\rho \mapsto (\langle00| +\langle11|)\rho(|00\rangle  + |11\rangle)$ \\
\hline
\end{tabular} 
\end{center}

\newpage

\begin{center}
{\bf (pure)  gates:}\\
\begin{tabular}{c|c|c|}
\hline
Notation:  & \had & \cnot \\ 
\hline
$\Dpure$:   & $\rho \mapsto \frac 12\left(\begin{array}{rr}
1 & 1\\
1 & -1
\end{array}\right)\rho \left(\begin{array}{rr}
1 & 1\\
1 & -1
\end{array}\right)$& $\rho\mapsto \cnotm \rho \cnotm$\\
\hline
\end{tabular} 
\end{center}

\vspace{3mm}

\begin{center}
{\bf CP maps:}\\
\begin{tabular}{c|c|c|c|}
\hline
Notation:  & \trace & \mix & \decohere \\ 
\hline
$\D$:   & \begin{tabular}{c} trace \\$\rho \mapsto tr(\rho)$ \end{tabular}  & 
\begin{tabular}{c} maximally \\ mixed  state\\ $1\mapsto \frac 12 \left(\begin{array}{rr}
1 & 0\\
0& 1
\end{array}\right)$\end{tabular} & 
\begin{tabular}{c} erase \\non-diagonal  elements\\$\rho \mapsto \langle 0|\rho |0\rangle +\langle 1|\rho |1\rangle $ \end{tabular}\\
\hline
\end{tabular} 
\end{center}

\vspace{3mm}

\begin{center}
{\bf (destructive) measurements:}\\
\begin{tabular}{c|c|c|cl}
\hline
Notation:  & 
 \begin{tikzpicture}[cat,scale=0.8]
\node (in) at (0,1.2) {};
\node (out) at (0,-0.8) {};
\ground{gr}{0.7,0.7}
\node [black vertex] (gdot) at (0,0) {};
\draw[thick,rounded corners=2pt] (gdot) -- (0.7,0.5)-- (gr);
\draw[thick] (gdot) -- (in);
\draw[thick] (gdot) -- (out);
\end{tikzpicture}
 & 
  \begin{tikzpicture}[cat,scale=0.8]
\node (in) at (0,1.2) {};
\node (out) at (0,-1.5) {};
\ground{gr}{0.7,0.7}
\node[h vertex] (h) at (0,-0.7) {};
\node [black vertex] (gdot) at (0,0) {};
\draw[thick,rounded corners=2pt] (gdot) -- (0.7,0.5)-- (gr);
\draw[thick] (gdot) -- (in);
\draw[thick] (gdot) -- (h) -- (out);
\end{tikzpicture}
&
 \begin{tikzpicture}[cat,scale=0.8]
\node (out) at (-1.4,2.4) {};
\node (out2) at (0,2.4) {};
\node[black vertex] (f) at (-1.4,0) {};
\node[black vertex] (g1) at (-1.4,1.4) {};
\node[black vertex] (g2) at (0,1.4) {};
\ground{gr1}{-2,2};
\ground{gr2}{0.6,2};
\node (inc) at (0,-1.4) {};
\node (inq) at (-1.4,-1.4) {};
\node[black vertex] (gdot) at (0,0) {};
\node[h vertex] (h) at (-0.7,0) {};
\node[h vertex] (h1) at (0,-0.7) {};
\node[h vertex] (h2) at (0,0.7) {};
\node[h vertex] (h3) at (-1.4,0.7) {};
\draw[thick,rounded corners=2pt] (gdot)  -- (h) -- (f);
\draw[thick] (inq)  -- (f) -- (h3) -- (out);
\draw[thick] (out2)-- (h2) -- (gdot) --(h1)-- (inc);
\draw[thick,rounded corners=2pt] (gr1)  -- (-2,1.8) -- (g1);
\draw[thick,rounded corners=2pt] (gr2)  -- (0.6,1.8) -- (g2);
\end{tikzpicture}
  \\ 
\hline
$\D$:   & Pauli $Z$ measurement &Pauli $X$ measurement & Bell-basis measurement \\
\hline
\end{tabular} 
\end{center}

\vspace{3mm}

\begin{center}
{\bf classically controlled operations:}\\
\begin{tabular}{c|c|c|}
\hline
Notation:  & 
\begin{tikzpicture}[cat,scale=0.8]
\node (out) at (-1,1.4) {};
\node[black vertex] (f) at (-1,0.5) {};
\node (inc) at (0,-0.8) {};
\node (inq) at (-1,-0.8) {};
\ground{gr}{0.4,0.7}
\node[black vertex] (gdot) at (0,0) {};
\node[h vertex] (h) at (-0.5,0.25) {};
\draw[thick,rounded corners=2pt] (gdot) -- (0.4,0.5)-- (gr); 
\draw[thick,rounded corners=5pt] (gdot)  -- (h) -- (f);
\draw[thick] (inq) -- (f);
\draw[thick] (out) -- (f);
\draw[thick] (gdot) -- (inc);
\end{tikzpicture}
 & 
\begin{tikzpicture}[cat,scale=0.8]
\node (out) at (-1,1.6) {};
\node[black vertex] (f) at (-1,0.5) {};
\node (inc) at (0,-0.8) {};
\node (inq) at (-1,-0.8) {};
\ground{gr}{0.4,0.7}
\node[black vertex] (gdot) at (0,0) {};
\node[h vertex] (h) at (-0.5,0.25) {};
\node[h vertex] (h1) at (-1,0) {};
\node[h vertex] (h2) at (-1,1) {};
\draw[thick,rounded corners=2pt] (gdot) -- (0.4,0.5)-- (gr); 
\draw[thick,rounded corners=5pt] (gdot)  -- (h) -- (f);
\draw[thick] (inq) -- (h1) -- (f);
\draw[thick] (out) --  (h2) -- (f);
\draw[thick] (gdot) -- (inc);
\end{tikzpicture}
  \\ 
\hline
$\D$:   & Pauli $Z$ correction & Pauli $X$ correction \\
\hline
\end{tabular} 
\end{center}

\section{Example protocols}\label{sec:examples}

In the statement of each proposition, we will specify protocols with explicit physical types, quantum channels being represented by full lines and classical channels being represented by dotted lines.  We use the symbol `$:\simeq$' for the passage of this specification to the interpretation within the diagrammatic calculus.

First we show that the teleportation protocol, by means of a Bell state and two classical channels, realizes a (perfect) quantum channel.

\begin{prop}[correctness of teleportation]
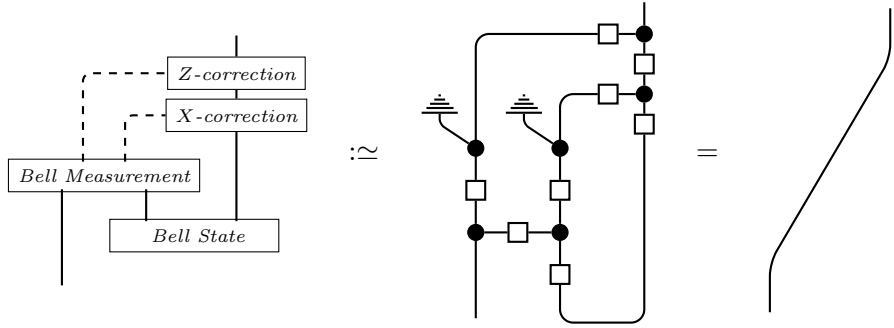
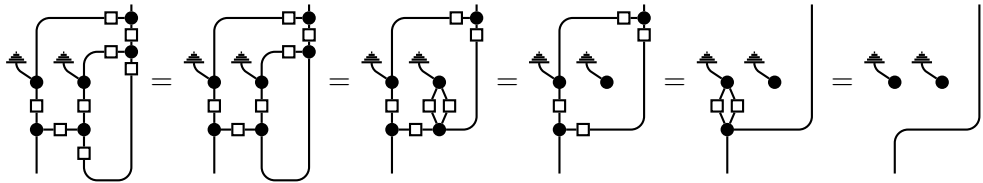
\[
\begin{tikzpicture}[cat,scale=0.8]
\node (in) at (-0.7,-1) {};
\node (out) at (2.2,3.5) {};
\node[box vertex] (b2) at (0,1){\tiny Bell Measurement};
\node[box vertex] (b1) at (1.5,0){\tiny \ \ \ \ Bell State\ \ \ \ };
\node[box vertex] (z) at (2.2,2){\tiny $X$-correction};
\node[box vertex] (x) at (2.2,2.7){\tiny $Z$-correction};
\draw[thick,rounded corners=8pt] (in) -- (-0.7,0.77);
\draw[thick,rounded corners=8pt] (0.7,0.77) -- (0.7,0.24);
\draw[thick,rounded corners=8pt] (out) -- (x) --(z) -- (2.2,0.24);
\draw[thick, dashed,rounded corners=3pt] (0.35,1.23) -- (0.35,2) -- (z);
\draw[thick, dashed, rounded corners=3pt] (-0.35,1.23) -- (-0.35,2.7) -- (x);
\end{tikzpicture}\ \ \ \
\centering{:\simeq}\ \ \ \
\begin{tikzpicture}[cat,scale=0.8]
\node (out) at (-1.4,2.4) {};
\node (out3) at (1.4,4) {};
\node[black vertex] (f) at (-1.4,0) {};
\node[black vertex] (g1) at (-1.4,1.4) {};
\node[black vertex] (g2) at (0,1.4) {};
\ground{gr1}{-2,2};
\ground{gr2}{-0.6,2};
\node (inq) at (-1.4,-1.6) {};
\node[black vertex] (gdot) at (0,0) {};
\node[h vertex] (h) at (-0.7,0) {};
\node[h vertex] (h1) at (0,-0.7) {};
\node[h vertex] (h2) at (0,0.7) {};
\node[h vertex] (h3) at (-1.4,0.7) {};
\node[h vertex] (hx1) at (1.4,1.8) {};
\node[black vertex] (gx) at (1.4,2.3) {};
\node[h vertex] (hx2) at (1.4,2.8) {};
\node[h vertex] (hx3) at (0.8,2.3) {};
\node[h vertex] (hz) at (0.8,3.3) {};
\node[black vertex] (gz) at (1.4,3.3) {};
\draw[thick,rounded corners=2pt] (gdot)  -- (h) -- (f);
\draw[thick,rounded corners=5pt] (inq)  -- (f) -- (h3) -- (-1.4,3.3) -- (hz) -- (gz);
\draw[thick,rounded corners=5pt]  (gx)-- (hx3) -- (0,2.3)-- (h2) -- (gdot) --(h1)-- (0,-1.5) -- (1.4,-1.5)--(hx1) -- (gx) -- (hx2) -- (gz)-- (out3);
\draw[thick,rounded corners=2pt] (gr1)  -- (-2,1.8) -- (g1);
\draw[thick,rounded corners=2pt] (gr2)  -- (-0.6,1.8) -- (g2);
\end{tikzpicture}
\ \ \ \
\centering{=}\ \ \ \ \begin{tikzpicture}[cat,scale=0.8]
\node (in) at (0,-1.4) {};
\node (out) at (2,4) {};
\draw[thick,rounded corners=5pt] (in) -- (0,-0.4)--(2,3)--(out);
\end{tikzpicture}
\]
\end{prop}
\begin{proof}
\[
\begin{tikzpicture}[cat,scale=0.45]
\node (out) at (-1.4,2.4) {};
\node (out3) at (1.4,4) {};
\node[black small vertex] (f) at (-1.4,0) {};
\node[black small vertex] (g1) at (-1.4,1.4) {};
\node[black small vertex] (g2) at (0,1.4) {};
\ground{gr1}{-2,2};
\ground{gr2}{-0.6,2};
\node (inq) at (-1.4,-1.6) {};
\node[black small vertex] (gdot) at (0,0) {};
\node[h small vertex] (h) at (-0.7,0) {};
\node[h small vertex] (h1) at (0,-0.7) {};
\node[h small vertex] (h2) at (0,0.7) {};
\node[h small vertex] (h3) at (-1.4,0.7) {};
\node[h small vertex] (hx1) at (1.4,1.8) {};
\node[black small vertex] (gx) at (1.4,2.3) {};
\node[h small vertex] (hx2) at (1.4,2.8) {};
\node[h small vertex] (hx3) at (0.8,2.3) {};
\node[h small vertex] (hz) at (0.8,3.3) {};
\node[black small vertex] (gz) at (1.4,3.3) {};
\draw[thick,rounded corners=2pt] (gdot)  -- (h) -- (f);
\draw[thick,rounded corners=5pt] (inq)  -- (f) -- (h3) -- (-1.4,3.3) -- (hz) -- (gz);
\draw[thick,rounded corners=5pt]  (gx)-- (hx3) -- (0,2.3)-- (h2) -- (gdot) --(h1)-- (0,-1.5) -- (1.4,-1.5)--(hx1) -- (gx) -- (hx2) -- (gz)-- (out3);
\draw[thick,rounded corners=2pt] (gr1)  -- (-2,1.8) -- (g1);
\draw[thick,rounded corners=2pt] (gr2)  -- (-0.6,1.8) -- (g2);
\end{tikzpicture}\
\centering{=}\
\begin{tikzpicture}[cat,scale=0.45]
\node (out) at (-1.4,2.4) {};
\node (out3) at (1.4,4) {};
\node[black small vertex] (f) at (-1.4,0) {};
\node[black small vertex] (g1) at (-1.4,1.4) {};
\node[black small vertex] (g2) at (0,1.4) {};
\ground{gr1}{-2,2};
\ground{gr2}{-0.6,2};
\node (inq) at (-1.4,-1.6) {};
\node[black small vertex] (gdot) at (0,0) {};
\node[h small vertex] (h) at (-0.7,0) {};
\node[h small vertex] (h2) at (0,0.7) {};
\node[h small vertex] (h3) at (-1.4,0.7) {};
\node[black small vertex] (gx) at (1.4,2.3) {};
\node[h small vertex] (hx2) at (1.4,2.8) {};
\node[h small vertex] (hx3) at (0.8,2.3) {};
\node[h small vertex] (hz) at (0.8,3.3) {};
\node[black small vertex] (gz) at (1.4,3.3) {};
\draw[thick,rounded corners=2pt] (gdot)  -- (h) -- (f);
\draw[thick,rounded corners=5pt] (inq)  -- (f) -- (h3) -- (-1.4,3.3) --(hz) --  (gz);
\draw[thick,rounded corners=5pt]  (gx)-- (hx3)--(0,2.3)-- (h2) -- (gdot) -- (0,-1.5) -- (1.4,-1.5) -- (gx) -- (hx2) -- (gz)-- (out3);
\draw[thick,rounded corners=2pt] (gr1)  -- (-2,1.8) -- (g1);
\draw[thick,rounded corners=2pt] (gr2)  -- (-0.6,1.8) -- (g2);
\end{tikzpicture}
\
\centering{=}\
\begin{tikzpicture}[cat,scale=0.45]
\node (out) at (-1.4,2.4) {};
\node (out3) at (1.1,4) {};
\node[black small vertex] (f) at (-1.4,0) {};
\node[black small vertex] (g1) at (-1.4,1.4) {};
\node[black small vertex] (g2) at (0,1.4) {};
\ground{gr1}{-2,2};
\ground{gr2}{-0.6,2};
\node (inq) at (-1.4,-1.6) {};
\node[black small vertex] (gdot) at (0,0) {};
\node[h small vertex] (h) at (-0.7,0) {};
\node[h small vertex] (h2) at (-0.3,0.7) {};
\node[h small vertex] (h3) at (-1.4,0.7) {};
\node[h small vertex] (hx2) at (1.1,2.8) {};
\node[h small vertex] (hx3) at (0.3,0.7) {};
\node[h small vertex] (hz) at (0.5,3.3) {};
\node[black small vertex] (gz) at (1.1,3.3) {};
\draw[thick,rounded corners=2pt] (gdot)  -- (h) -- (f);
\draw[thick,rounded corners=5pt] (inq)  -- (f) -- (h3) -- (-1.4,3.3) --(hz)-- (gz);
\draw[thick,rounded corners=5pt]  (gdot) --(hx3) -- (g2)-- (h2) -- (gdot) -- (1.1,0)-- (hx2) -- (gz)-- (out3);
\draw[thick,rounded corners=2pt] (gr1)  -- (-2,1.8) -- (g1);
\draw[thick,rounded corners=2pt] (gr2)  -- (-0.6,1.8) -- (g2);
\end{tikzpicture}
\
\centering{=}\
\begin{tikzpicture}[cat,scale=0.45]
\node (out) at (-1.4,2.4) {};
\node (out3) at (1.1,4) {};
\node[black small vertex] (f) at (-1.4,0) {};
\node[black small vertex] (g1) at (-1.4,1.4) {};
\node[black small vertex] (g2) at (0,1.4) {};
\ground{gr1}{-2,2};
\ground{gr2}{-0.6,2};
\node (inq) at (-1.4,-1.6) {};
\node[h small vertex] (h) at (-0.7,0) {};
\node[h small vertex] (h3) at (-1.4,0.7) {};
\node[h small vertex] (hx2) at (1.1,2.8) {};
\node[h small vertex] (hz) at (0.5,3.3) {};
\node[black small vertex] (gz) at (1.1,3.3) {};
\draw[thick,rounded corners=2pt] (h) -- (f);
\draw[thick,rounded corners=5pt] (inq)  -- (f) -- (h3) -- (-1.4,3.3) --(hz)-- (gz);
\draw[thick,rounded corners=5pt]  (h) -- (1.1,0)-- (hx2) -- (gz)-- (out3);
\draw[thick,rounded corners=2pt] (gr1)  -- (-2,1.8) -- (g1);
\draw[thick,rounded corners=2pt] (gr2)  -- (-0.6,1.8) -- (g2);
\end{tikzpicture}
\
\centering{=}\
\begin{tikzpicture}[cat,scale=0.45]
\node (out) at (-1.4,2.4) {};
\node (out3) at (1.1,4) {};
\node[black small vertex] (f) at (-1.4,0) {};
\node[black small vertex] (g1) at (-1.4,1.4) {};
\node[black small vertex] (g2) at (0,1.4) {};
\ground{gr1}{-2,2};
\ground{gr2}{-0.6,2};
\node (inq) at (-1.4,-1.6) {};
\node[h small vertex] (h3) at (-1.7,0.7) {};
\node[h small vertex] (hz) at (-1.1,0.7) {};
\draw[thick,rounded corners=5pt] (inq)  -- (f) -- (h3) -- (g1) --(hz)-- (f);
\draw[thick,rounded corners=5pt]  (f) -- (1.1,0)-- (out3);
\draw[thick,rounded corners=2pt] (gr1)  -- (-2,1.8) -- (g1);
\draw[thick,rounded corners=2pt] (gr2)  -- (-0.6,1.8) -- (g2);
\end{tikzpicture}\
\centering{=}\
\begin{tikzpicture}[cat,scale=0.45]
\node (out) at (-1.4,2.4) {};
\node (out3) at (1.1,4) {};
\node[black small vertex] (g1) at (-1.4,1.4) {};
\node[black small vertex] (g2) at (0,1.4) {};
\ground{gr1}{-2,2};
\ground{gr2}{-0.6,2};
\node (inq) at (-1.4,-1.6) {};
\draw[thick,rounded corners=5pt] (inq)  -- (-1.4,0) -- (1.1,0)-- (out3);
\draw[thick,rounded corners=2pt] (gr1)  -- (-2,1.8) -- (g1);
\draw[thick,rounded corners=2pt] (gr2)  -- (-0.6,1.8) -- (g2);
\end{tikzpicture}
\]
\end{proof}

We show that the state transfer \cite{Perdrix} protocol, by means of $2$-qubit unitary tranformation and a local measurement, realizes a (perfect) quantum channel.

\begin{prop}[correctness of state transfer]
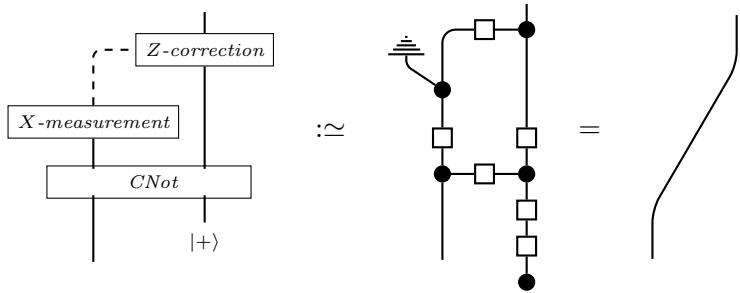
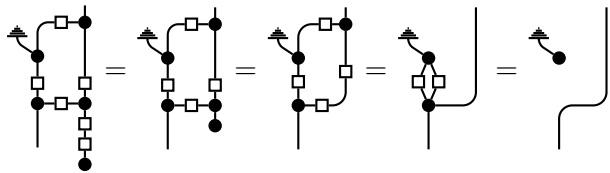
\[
\begin{tikzpicture}[cat,scale=0.8]
\node (in) at (0.35,-1) {};
\node (out) at (2.2,3.5) {};
\node (p) at (2.2, -0.5) {\tiny $|+\rangle$};
\node[box vertex] (b2) at (1.275,0.5){\tiny\ \ \ \ \ \ \ \ \ CNot\ \ \ \ \ \ \ \ \ };
\node[box vertex] (z) at (2.2,2.7){\tiny $Z$-correction};
\node[box vertex] (x) at (0.35,1.5){\tiny $X$-measurement};
\draw[thick,rounded corners=8pt] (in) -- (0.35,0.28);
\draw[thick,rounded corners=8pt] (2.2,0.28) -- (p);
\draw[thick,rounded corners=8pt] (0.35,0.72) -- (x);
\draw[thick,rounded corners=8pt] (out) --(z) -- (2.2,0.72);
\draw[thick, dashed,rounded corners=3pt] (x) -- (0.35,2.7) -- (z);
\end{tikzpicture}\ \ \ \
\centering{:\simeq}\ \ \ \
\begin{tikzpicture}[cat,scale=0.8]
\node (out) at (-1.4,2.4) {};
\node (out3) at (0,3) {};
\node[black vertex] (f) at (-1.4,0) {};
\node[black vertex] (g1) at (-1.4,1.4) {};
\node[black vertex] (g2) at (0,-1.8) {};
\ground{gr1}{-2,2};
\node (inq) at (-1.4,-1.6) {};
\node[black vertex] (gdot) at (0,0) {};
\node[h vertex] (h) at (-0.7,0) {};
\node[h vertex] (h1) at (0,-0.6) {};
\node[h vertex] (h4) at (0,-1.2) {};
\node[h vertex] (h2) at (0,0.6) {};
\node[h vertex] (h3) at (-1.4,0.6) {};
\node[h vertex] (hz) at (-0.7,2.4) {};
\node[black vertex] (gz) at (0,2.4) {};
\draw[thick,rounded corners=2pt] (gdot)  -- (h) -- (f);
\draw[thick,rounded corners=5pt] (inq)  -- (f) -- (h3) -- (-1.4,2.4) -- (hz) -- (gz);
\draw[thick,rounded corners=5pt]   (out3) -- (h2) -- (gdot) --(h1);
\draw[thick,rounded corners=2pt] (gr1)  -- (-2,1.8) -- (g1);
\draw[thick,rounded corners=2pt] (h1)  -- (h4) -- (g2);
\end{tikzpicture}
\ \ \ \
\centering{=}\ \ \ \ \begin{tikzpicture}[cat,scale=0.8]
\node (in) at (0,-1.4) {};
\node (out) at (1.4,3) {};
\draw[thick,rounded corners=5pt] (in) -- (0,-0.4)--(1.4,2)--(out);
\end{tikzpicture}
\]
\end{prop}
\begin{proof}
\[
\begin{tikzpicture}[cat,scale=0.45]
\node (out) at (-1.4,2.4) {};
\node (out3) at (0,3.2) {};
\node[black small vertex] (f) at (-1.4,0) {};
\node[black small vertex] (g1) at (-1.4,1.4) {};
\node[black small vertex] (g2) at (0,-1.8) {};
\ground{gr1}{-2,2};
\node (inq) at (-1.4,-1.6) {};
\node[black small vertex] (gdot) at (0,0) {};
\node[h small vertex] (h) at (-0.7,0) {};
\node[h small vertex] (h1) at (0,-0.6) {};
\node[h small vertex] (h4) at (0,-1.2) {};
\node[h small vertex] (h2) at (0,0.6) {};
\node[h small vertex] (h3) at (-1.4,0.6) {};
\node[h small vertex] (hz) at (-0.7,2.4) {};
\node[black small vertex] (gz) at (0,2.4) {};
\draw[thick,rounded corners=2pt] (gdot)  -- (h) -- (f);
\draw[thick,rounded corners=4pt] (inq)  -- (f) -- (h3) -- (-1.4,2.4) -- (hz) -- (gz);
\draw[thick,rounded corners=4pt]   (out3) -- (h2) -- (gdot) --(h1);
\draw[thick,rounded corners=2pt] (gr1)  -- (-2,1.8) -- (g1);
\draw[thick,rounded corners=2pt] (h1)  -- (h4) -- (g2);
\end{tikzpicture}\
\centering{=}\
\begin{tikzpicture}[cat,scale=0.45]
\node (out) at (-1.4,2.4) {};
\node (out3) at (0,3.2) {};
\node[black small vertex] (f) at (-1.4,0) {};
\node[black small vertex] (g1) at (-1.4,1.4) {};
\node[black small vertex] (g2) at (0,-0.6) {};
\ground{gr1}{-2,2};
\node (inq) at (-1.4,-1.6) {};
\node[black small vertex] (gdot) at (0,0) {};
\node[h small vertex] (h) at (-0.7,0) {};
\node[h small vertex] (h2) at (0,0.6) {};
\node[h small vertex] (h3) at (-1.4,0.6) {};
\node[h small vertex] (hz) at (-0.7,2.4) {};
\node[black small vertex] (gz) at (0,2.4) {};
\draw[thick,rounded corners=2pt] (gdot)  -- (h) -- (f);
\draw[thick,rounded corners=4pt] (inq)  -- (f) -- (h3) -- (-1.4,2.4) -- (hz) -- (gz);
\draw[thick,rounded corners=4pt]   (out3) -- (h2) -- (gdot) --(g2);
\draw[thick,rounded corners=2pt] (gr1)  -- (-2,1.8) -- (g1);
\end{tikzpicture}\
\centering{=}\
\begin{tikzpicture}[cat,scale=0.45]
\node (out) at (-1.4,2.4) {};
\node (out3) at (0,3.2) {};
\node[black small vertex] (f) at (-1.4,0) {};
\node[black small vertex] (g1) at (-1.4,1.4) {};
\ground{gr1}{-2,2};
\node (inq) at (-1.4,-1.6) {};
\node[h small vertex] (h) at (-0.7,0) {};
\node[h small vertex] (h3) at (-1.4,0.7) {};
\node[h small vertex] (hx2) at (0,1) {};
\node[h small vertex] (hz) at (-0.6,2.4) {};
\node[black small vertex] (gz) at (0,2.4) {};
\draw[thick,rounded corners=2pt] (h) -- (f);
\draw[thick,rounded corners=5pt] (inq)  -- (f) -- (h3) -- (-1.4,2.4) --(hz)-- (gz);
\draw[thick,rounded corners=5pt]  (h) -- (0,0)-- (hx2) -- (gz)-- (out3);
\draw[thick,rounded corners=2pt] (gr1)  -- (-2,1.8) -- (g1);
\end{tikzpicture}
\
\centering{=}\
\begin{tikzpicture}[cat,scale=0.45]
\node (out) at (-1.4,2.4) {};
\node (out3) at (0,3.2) {};
\node[black small vertex] (f) at (-1.4,0) {};
\node[black small vertex] (g1) at (-1.4,1.4) {};
\ground{gr1}{-2,2};
\node (inq) at (-1.4,-1.6) {};
\node[h small vertex] (h3) at (-1.7,0.7) {};
\node[h small vertex] (hz) at (-1.1,0.7) {};
\draw[thick,rounded corners=5pt] (inq)  -- (f) -- (h3) -- (g1) --(hz)-- (f);
\draw[thick,rounded corners=5pt]  (f) -- (0,0)-- (out3);
\draw[thick,rounded corners=2pt] (gr1)  -- (-2,1.8) -- (g1);
\end{tikzpicture}\
\centering{=}\
\begin{tikzpicture}[cat,scale=0.45]
\node (out) at (-1.4,2.4) {};
\node (out3) at (0,3.2) {};
\node[black small vertex] (g1) at (-1.4,1.4) {};
\ground{gr1}{-2,2};
\node (inq) at (-1.4,-1.6) {};
\draw[thick,rounded corners=5pt] (inq)  -- (-1.4,0) -- (0,0)-- (out3);
\draw[thick,rounded corners=2pt] (gr1)  -- (-2,1.8) -- (g1);
\end{tikzpicture}
\]
\end{proof}

Now we show that the dense coding protocol, by means of a Bell state and a quantum channel, realizes two classical channels.

\begin{prop}[correctness of dense coding]
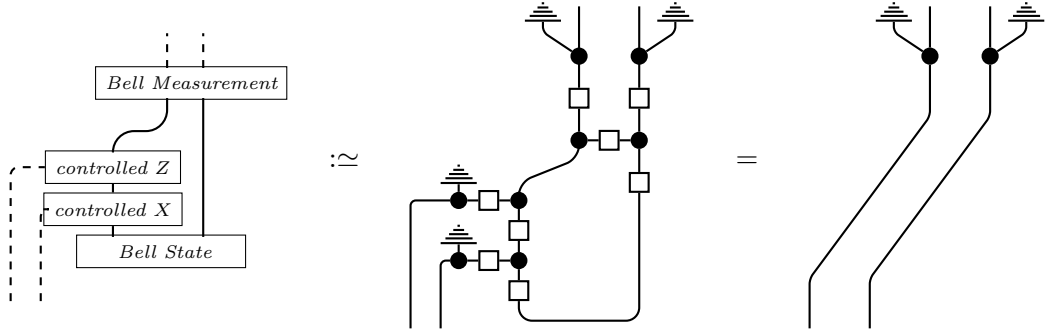
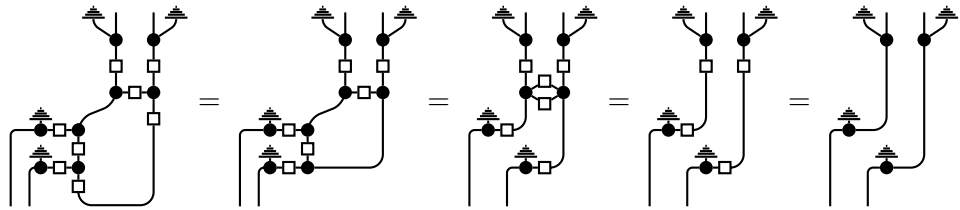
\[
\begin{tikzpicture}[cat,scale=0.8]
\node (inc1) at (-0.5,-1) {};
\node (inc2) at (-1,-1) {};
\node (out2) at (2.2,3.8) {};
\node (out1) at (1.6,3.8) {};
\node[box vertex] (b2) at (2,2.8){\tiny Bell Measurement};
\node[box vertex] (b1) at (1.5,0){\tiny \ \ \ \ Bell State\ \ \ \ };
\node[box vertex] (x) at (0.7,0.7){\tiny controlled $X$};
\node[box vertex] (z) at (0.7,1.4){\tiny controlled $Z$};
\draw[thick, dashed,rounded corners=3pt] (inc1) -- (-0.5,0.7) -- (x);
\draw[thick, dashed,rounded corners=3pt] (inc2) -- (-1,1.4) -- (z);
\draw[thick,rounded corners=8pt] (0.7,0.24) -- (x) --(z) --(0.7,2) -- (1.6,2) -- (1.6,2.55) ;
\draw[thick,rounded corners=8pt] (2.2,2.55) -- (2.2,0.24);
\draw[thick, dashed,rounded corners=3pt] (2.2,3.05) -- (out2) ;
\draw[thick, dashed,rounded corners=3pt] (1.6,3.05) -- (out1) ;
\end{tikzpicture}\ \ \ \
\centering{:\simeq}\ \ \ \
\begin{tikzpicture}[cat,scale=0.8]
\node (out) at (-1,2.4) {};
\node (out2) at (0,2.4) {};
\node[black vertex] (f) at (-1,0) {};
\node[black vertex] (g1) at (-1,1.4) {};
\node[black vertex] (g2) at (0,1.4) {};
\node[black vertex] (gz) at (-2,-1) {};
\node[black vertex] (gx) at (-2,-2) {};
\node[h vertex] (hx1) at (-2,-1.5) {};
\node[h vertex] (hx2) at (-2,-2.5) {};
\node[h vertex] (hx) at (-2.5,-2) {};
\node[h vertex] (hz) at (-2.5,-1) {};
\ground{gr1}{-1.6,2};
\ground{gr2}{0.6,2};
\node (inc2) at (-3.3,-3.3) {};
\node (inc1) at (-3.8,-3.3) {};
\node[black vertex] (gdot) at (0,0) {};
\node[h vertex] (h) at (-0.5,0) {};
\node[h vertex] (h1) at (0,-0.7) {};
\node[h vertex] (h2) at (0,0.7) {};
\node[h vertex] (h3) at (-1,0.7) {};
\ground{gr3}{-3,-1.7};
\ground{gr4}{-3,-0.7};
\node[black vertex] (gc3) at (-3,-2) {};
\node[black vertex] (gc4) at (-3,-1) {};
\draw[thick,rounded corners=2pt] (gdot)  -- (h) -- (f);
\draw[thick,rounded corners=5pt] (h1) -- (0,-3)--(-2,-3) --(hx2) -- (gx) --(hx1) -- (gz)-- (-2,-0.7)-- (-1,-0.3)  -- (f) -- (h3) -- (out);
\draw[thick] (out2)-- (h2) -- (gdot) --(h1);
\draw[thick,rounded corners=2pt] (gr1)  -- (-1.6,1.8) -- (g1);
\draw[thick,rounded corners=2pt] (gr2)  -- (0.6,1.8) -- (g2);
\draw[thick,rounded corners=2pt] (gc3)   -- (gr3);
\draw[thick,rounded corners=2pt] (gc4)   -- (gr4);
\draw[thick,rounded corners=2pt] (gx)  -- (hx) -- (gc3) -- (-3.3,-2) -- (inc2);
\draw[thick,rounded corners=2pt] (gz)  -- (hz) -- (gc4) -- (-3.8,-1) -- (inc1);
\end{tikzpicture}
\ \ \ \
\centering{=}\ \ \ \
\begin{tikzpicture}[cat,scale=0.8]
\node (out) at (-1,2.4) {};
\node (out2) at (0,2.4) {};
\node[black vertex] (g1) at (-1,1.4) {};
\node[black vertex] (g2) at (0,1.4) {};
\ground{gr1}{-1.6,2};
\ground{gr2}{0.6,2};
\node (inc2) at (-2,-3.3) {};
\node (inc1) at (-3,-3.3) {};
\draw[thick,rounded corners=2pt] (gr1)  -- (-1.6,1.8) -- (g1);
\draw[thick,rounded corners=2pt] (gr2)  -- (0.6,1.8) -- (g2);
\draw[thick,rounded corners=2pt] (inc1) --(-3,-2.3)--(-1,0.4) --(g1) --(out);
\draw[thick,rounded corners=2pt] (inc2) --(-2,-2.3)--(0,0.4) --(g2) --(out2);
\end{tikzpicture}
\]
\end{prop}
\begin{proof}
\[\begin{tikzpicture}[cat,scale=0.5]
\node (out) at (-1,2.4) {};
\node (out2) at (0,2.4) {};
\node[black small vertex] (f) at (-1,0) {};
\node[black small vertex] (g1) at (-1,1.4) {};
\node[black small vertex] (g2) at (0,1.4) {};
\node[black small vertex] (gz) at (-2,-1) {};
\node[black small vertex] (gx) at (-2,-2) {};
\node[h small vertex] (hx1) at (-2,-1.5) {};
\node[h small vertex] (hx2) at (-2,-2.5) {};
\node[h small vertex] (hx) at (-2.5,-2) {};
\node[h small vertex] (hz) at (-2.5,-1) {};
\ground{gr1}{-1.6,2};
\ground{gr2}{0.6,2};
\node (inc2) at (-3.3,-3.3) {};
\node (inc1) at (-3.8,-3.3) {};
\node[black small vertex] (gdot) at (0,0) {};
\node[h small vertex] (h) at (-0.5,0) {};
\node[h small vertex] (h1) at (0,-0.7) {};
\node[h small vertex] (h2) at (0,0.7) {};
\node[h small vertex] (h3) at (-1,0.7) {};
\ground{gr3}{-3,-1.7};
\ground{gr4}{-3,-0.7};
\node[black small vertex] (gc3) at (-3,-2) {};
\node[black small vertex] (gc4) at (-3,-1) {};
\draw[thick,rounded corners=2pt] (gdot)  -- (h) -- (f);
\draw[thick,rounded corners=5pt] (h1) -- (0,-3)--(-2,-3) --(hx2) -- (gx) --(hx1) -- (gz)-- (-2,-0.7)-- (-1,-0.3)  -- (f) -- (h3) -- (out);
\draw[thick] (out2)-- (h2) -- (gdot) --(h1);
\draw[thick,rounded corners=2pt] (gr1)  -- (-1.6,1.8) -- (g1);
\draw[thick,rounded corners=2pt] (gr2)  -- (0.6,1.8) -- (g2);
\draw[thick,rounded corners=2pt] (gc3)   -- (gr3);
\draw[thick,rounded corners=2pt] (gc4)   -- (gr4);
\draw[thick,rounded corners=2pt] (gx)  -- (hx) -- (gc3) -- (-3.3,-2) -- (inc2);
\draw[thick,rounded corners=2pt] (gz)  -- (hz) -- (gc4) -- (-3.8,-1) -- (inc1);
\end{tikzpicture}\ \centering{=}\
\begin{tikzpicture}[cat,scale=0.5]
\node (out) at (-1,2.4) {};
\node (out2) at (0,2.4) {};
\node[black small vertex] (f) at (-1,0) {};
\node[black small vertex] (g1) at (-1,1.4) {};
\node[black small vertex] (g2) at (0,1.4) {};
\node[black small vertex] (gz) at (-2,-1) {};
\node[black small vertex] (gx) at (-2,-2) {};
\node[h small vertex] (hx1) at (-2,-1.5) {};
\node[h small vertex] (hx) at (-2.5,-2) {};
\node[h small vertex] (hz) at (-2.5,-1) {};
\ground{gr1}{-1.6,2};
\ground{gr2}{0.6,2};
\node (inc2) at (-3.3,-3.3) {};
\node (inc1) at (-3.8,-3.3) {};
\node[black small vertex] (gdot) at (0,0) {};
\node[h small vertex] (h) at (-0.5,0) {};
\node[h small vertex] (h2) at (0,0.7) {};
\node[h small vertex] (h3) at (-1,0.7) {};
\ground{gr3}{-3,-1.7};
\ground{gr4}{-3,-0.7};
\node[black small vertex] (gc3) at (-3,-2) {};
\node[black small vertex] (gc4) at (-3,-1) {};
\draw[thick,rounded corners=2pt] (gdot)  -- (h) -- (f);
\draw[thick,rounded corners=5pt] (gdot) -- (0,-2) -- (gx) --(hx1) -- (gz)-- (-2,-0.7)-- (-1,-0.3)  -- (f) -- (h3) -- (out);
\draw[thick] (out2)-- (h2) -- (gdot) ;
\draw[thick,rounded corners=2pt] (gr1)  -- (-1.6,1.8) -- (g1);
\draw[thick,rounded corners=2pt] (gr2)  -- (0.6,1.8) -- (g2);
\draw[thick,rounded corners=2pt] (gc3)   -- (gr3);
\draw[thick,rounded corners=2pt] (gc4)   -- (gr4);
\draw[thick,rounded corners=2pt] (gx)  -- (hx) -- (gc3) -- (-3.3,-2) -- (inc2);
\draw[thick,rounded corners=2pt] (gz)  -- (hz) -- (gc4) -- (-3.8,-1) -- (inc1);
\end{tikzpicture}
\ \centering{=}\
\begin{tikzpicture}[cat,scale=0.5]
\node (out) at (-1,2.4) {};
\node (out2) at (0,2.4) {};
\node[black small vertex] (f) at (-1,0) {};
\node[black small vertex] (g1) at (-1,1.4) {};
\node[black small vertex] (g2) at (0,1.4) {};
\node[h small vertex] (hx1) at (-0.5,-0.3) {};
\node[h small vertex] (hx) at (-0.5,-2) {};
\node[h small vertex] (hz) at (-1.5,-1) {};
\ground{gr1}{-1.6,2};
\ground{gr2}{0.6,2};
\node (inc2) at (-1.5,-3.3) {};
\node (inc1) at (-2.5,-3.3) {};
\node[black small vertex] (gdot) at (0,0) {};
\node[h small vertex] (h) at (-0.5,0.3) {};
\node[h small vertex] (h2) at (0,0.7) {};
\node[h small vertex] (h3) at (-1,0.7) {};
\ground{gr3}{-1,-1.7};
\ground{gr4}{-2,-0.7};
\node[black small vertex] (gc3) at (-1,-2) {};
\node[black small vertex] (gc4) at (-2,-1) {};
\draw[thick,rounded corners=2pt] (gdot)  -- (h) -- (f);
\draw[thick,rounded corners=2pt] (gdot)  -- (hx1) -- (f);
\draw[thick,rounded corners=5pt] (gdot) -- (0,-2) -- (hx);
\draw[thick,rounded corners=5pt]  (hz)-- (-1,-1)  -- (f) -- (h3) -- (out);
\draw[thick] (out2)-- (h2) -- (gdot) ;
\draw[thick,rounded corners=2pt] (gr1)  -- (-1.6,1.8) -- (g1);
\draw[thick,rounded corners=2pt] (gr2)  -- (0.6,1.8) -- (g2);
\draw[thick,rounded corners=2pt] (gc3)   -- (gr3);
\draw[thick,rounded corners=2pt] (gc4)   -- (gr4);
\draw[thick,rounded corners=2pt] (hx) -- (gc3) -- (-1.5,-2) -- (inc2);
\draw[thick,rounded corners=2pt]  (hz) -- (gc4) -- (-2.5,-1) -- (inc1);
\end{tikzpicture}
\ \centering{=}\
\begin{tikzpicture}[cat,scale=0.5]
\node (out) at (-1,2.4) {};
\node (out2) at (0,2.4) {};
\node[black small vertex] (g1) at (-1,1.4) {};
\node[black small vertex] (g2) at (0,1.4) {};
\node[h small vertex] (hx) at (-0.5,-2) {};
\node[h small vertex] (hz) at (-1.5,-1) {};
\ground{gr1}{-1.6,2};
\ground{gr2}{0.6,2};
\node (inc2) at (-1.5,-3.3) {};
\node (inc1) at (-2.5,-3.3) {};
\node[h small vertex] (h2) at (0,0.7) {};
\node[h small vertex] (h3) at (-1,0.7) {};
\ground{gr3}{-1,-1.7};
\ground{gr4}{-2,-0.7};
\node[black small vertex] (gc3) at (-1,-2) {};
\node[black small vertex] (gc4) at (-2,-1) {};
\draw[thick,rounded corners=5pt] (h2) -- (0,-2) -- (hx);
\draw[thick,rounded corners=5pt]  (hz)-- (-1,-1)   -- (h3) -- (out);
\draw[thick] (out2)-- (h2) ;
\draw[thick,rounded corners=2pt] (gr1)  -- (-1.6,1.8) -- (g1);
\draw[thick,rounded corners=2pt] (gr2)  -- (0.6,1.8) -- (g2);
\draw[thick,rounded corners=2pt] (gc3)   -- (gr3);
\draw[thick,rounded corners=2pt] (gc4)   -- (gr4);
\draw[thick,rounded corners=2pt] (hx) -- (gc3) -- (-1.5,-2) -- (inc2);
\draw[thick,rounded corners=2pt]  (hz) -- (gc4) -- (-2.5,-1) -- (inc1);
\end{tikzpicture}
\ \centering{=}\
\begin{tikzpicture}[cat,scale=0.5]
\node (out) at (-1,2.4) {};
\node (out2) at (0,2.4) {};
\node[black small vertex] (g1) at (-1,1.4) {};
\node[black small vertex] (g2) at (0,1.4) {};
\ground{gr1}{-1.6,2};
\ground{gr2}{0.6,2};
\node (inc2) at (-1.5,-3.3) {};
\node (inc1) at (-2.5,-3.3) {};
\ground{gr3}{-1,-1.7};
\ground{gr4}{-2,-0.7};
\node[black small vertex] (gc3) at (-1,-2) {};
\node[black small vertex] (gc4) at (-2,-1) {};
\draw[thick,rounded corners=5pt] (out2) -- (0,-2) -- (gc3);
\draw[thick,rounded corners=5pt]  (gc4)-- (-1,-1)  -- (out);
\draw[thick,rounded corners=2pt] (gr1)  -- (-1.6,1.8) -- (g1);
\draw[thick,rounded corners=2pt] (gr2)  -- (0.6,1.8) -- (g2);
\draw[thick,rounded corners=2pt] (gc3)   -- (gr3);
\draw[thick,rounded corners=2pt] (gc4)   -- (gr4);
\draw[thick,rounded corners=2pt] (gc3) -- (-1.5,-2) -- (inc2);
\draw[thick,rounded corners=2pt]   (gc4) -- (-2.5,-1) -- (inc1);
\end{tikzpicture}
\]
\end{proof} 

A diagrammatic presentation of a quantum key exchange protocols can be found  in \cite{CWWWZ}.  Here we provide a simplified presentation  by relying on the notion of environment. We restrict ourselves to  four representative cases.

\begin{prop}[correctness of BB84 and Ekert 91 key exchange]\em\ 
\bit
\item Alice and Bob  choose the same measurement in Ekert 91:
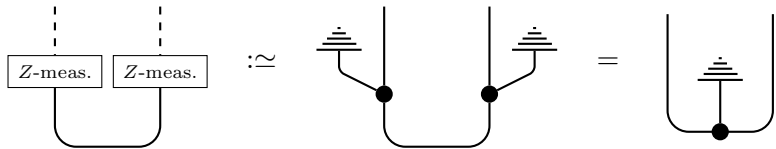
\[
\begin{tikzpicture}[cat,scale=1]
\node (o2) at (0,2) {};
\node (o1) at (1.4,2) {};
\node[box vertex] (b2) at (0,1){\tiny $Z$-meas.}; 
\node[box vertex] (b1) at (1.4,1){\tiny $Z$-meas.};
\draw[thick, dashed,rounded corners=8pt] (o1) -- (b1);
\draw[thick,rounded corners=8pt]  (b1) -- (1.4,0) -- (0,0) -- (b2);
\draw[thick, dashed,rounded corners=8pt] (o2) -- (b2);
\end{tikzpicture}\ \ \ \
\centering{:\simeq}\ \ \ \ \begin{tikzpicture}[cat,scale=1]
\node (o2) at (0,2) {};
\node (o1) at (1.4,2) {};
\node[black vertex] (b2) at (0,0.7){};
\node[black vertex] (b1) at (1.4,0.7){};
\ground{gr2}{-0.6,1.3};
\ground{gr1}{2,1.3};
\draw[thick,rounded corners=8pt] (o1) -- (b1)-- (1.4,0) -- (0,0) -- (b2) -- (o2);
\draw[thick,rounded corners=2pt] (b1) -- (2,1)-- (gr1);
\draw[thick,rounded corners=2pt] (b2) -- (-0.6,1)-- (gr2);
\end{tikzpicture}\ \ \ \ 
\centering{=}\ \ \ \ \begin{tikzpicture}[cat,scale=1]
\node (o2) at (0,1.7) {};
\node (o1) at (1.4,1.7) {};
\node[black vertex] (b) at (0.7,0){};
\ground{gr}{0.7,0.7};
\draw[thick,rounded corners=8pt] (o1) -- (1.4,0) -- (0,0)  -- (o2);
\draw[thick,rounded corners=2pt] (b) -- (gr);
\end{tikzpicture}\]
i.e.~Alice and Bob share the same classical data.


\item Alice and Bob  choose a different measurement in Ekert 91:
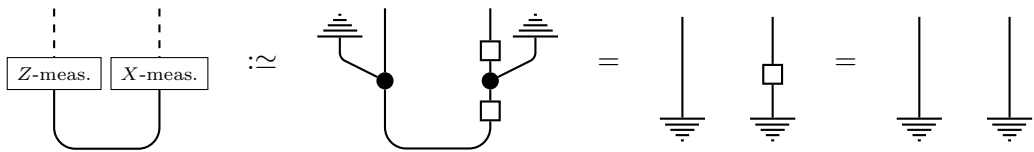
\[\begin{tikzpicture}[cat,scale=1]
\node (o2) at (0,2) {};
\node (o1) at (1.4,2) {};
\node[box vertex] (b2) at (0,1){\tiny $Z$-meas.};
\node[box vertex] (b1) at (1.4,1){\tiny $X$-meas.};
\draw[thick, dashed,rounded corners=8pt] (o1) -- (b1);
\draw[thick,rounded corners=8pt]  (b1) -- (1.4,0) -- (0,0) -- (b2);
\draw[thick, dashed,rounded corners=8pt] (o2) -- (b2);
\end{tikzpicture}\ \ \ \
\centering{:\simeq}\ \ \ \ \begin{tikzpicture}[cat,scale=1]
\node (o2) at (0,1.8) {};
\node (o1) at (1.4,1.8) {};
\node[black vertex] (b2) at (0,0.7){};
\node[black vertex] (b1) at (1.4,0.7){};
\node[h vertex](h1) at (1.4,0.3) {};
\node[h vertex](h2) at (1.4,1.1) {};
\ground{gr2}{-0.6,1.3};
\ground{gr1}{2,1.3};
\draw[thick,rounded corners=8pt] (o1) --(h2) -- (b1)-- (h1) -- (1.4,-0.2) -- (0,-0.20) -- (b2) -- (o2);
\draw[thick,rounded corners=2pt] (b1) -- (2,1)-- (gr1);
\draw[thick,rounded corners=2pt] (b2) -- (-0.6,1)-- (gr2);
\end{tikzpicture}\ \ \ \ 
\centering{=}\ \ \ \
\begin{tikzpicture}[cat,scale=1]
\node (o2) at (0,2) {};
\node (o1) at (1.2,2) {};
\maxmixed{gr1}{1.2,0.5}
\node[h vertex](h2) at (1.2,1.1) {};
\maxmixed{gr2}{0,0.5}
\draw[thick,rounded corners=8pt] (o1) -- (h2) -- (gr1);
\draw[thick,rounded corners=2pt] (o2) -- (gr2);
\end{tikzpicture}\ \ \ \
\centering{=}\ \ \ \
\begin{tikzpicture}[cat,scale=1]
\node (o2) at (0,2) {};
\node (o1) at (1.2,2) {};
\maxmixed{gr1}{1.2,0.5}
\maxmixed{gr2}{0,0.5}
\draw[thick,rounded corners=8pt] (o1) -- (gr1);
\draw[thick,rounded corners=2pt] (o2) -- (gr2);
\end{tikzpicture}
\]
i.e.~Alice's and Bob's data is not correlated.
\item Eve chooses the same  measurement as Alice and Bob in BB84:
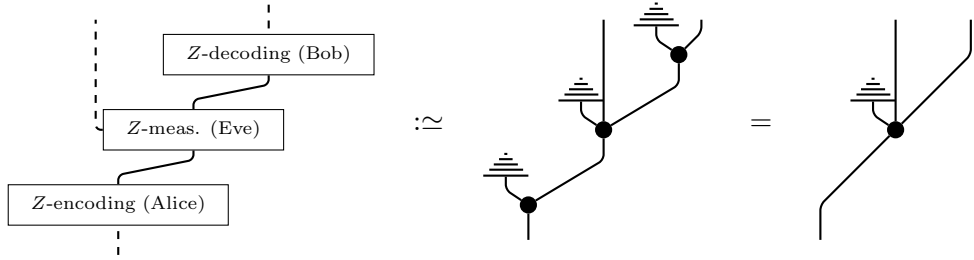
\[\begin{tikzpicture}[cat,scale=1]
\node (i) at (0,-0.8) {};
\node (o) at (2,2.8) {};
\node (oc) at (-0.3,2.6) {};
\node[box vertex] (m1) at (0,0){\tiny \begin{tabular}{c}$Z$-encoding (Alice)\end{tabular}};
\node[box vertex] (m2) at (1,1){\tiny \begin{tabular}{c}$Z$-meas. (Eve)\end{tabular}};
\node[box vertex] (m3) at (2,2){\tiny \begin{tabular}{c}$Z$-decoding (Bob)\end{tabular}};
\draw[thick, dashed,rounded corners=2pt] (i) -- (m1);
\draw[thick, dashed,rounded corners=2pt] (o) -- (m3);
\draw[thick,rounded corners=2pt] (m1)   --(0,0.4)--(1,0.6)-- (m2) --(1,1.4)--(2,1.6) --(m3);
\draw[thick, dashed,rounded corners=2pt] (m2)--(-0.3,1)--(oc);
\end{tikzpicture}
\ \ \ \
\centering{:\simeq}\ \ \ \ \begin{tikzpicture}[cat,scale=1]
\node (i) at (0,-0.6) {};
\node (o) at (2.3,2.6) {};
\node (oc) at (1,2.6) {};
\node[black vertex] (b1) at (0,0){};
\node[black vertex] (b2) at (1,1){};
\node[black vertex] (b3) at (2,2){};
\ground{gr1}{-0.3,0.4};
\ground{gr2}{0.7,1.4};
\ground{gr3}{1.7,2.4};
\draw[thick,rounded corners=2pt] (i) -- (b1) --(1,0.6)-- (b2)--(2,1.6)-- (b3) -- (2.3,2.3)--(o);
\draw[thick,rounded corners=2pt] (gr1)--(-0.3,0.2)--(b1);
\draw[thick,rounded corners=2pt] (gr2)--(0.7,1.2)--(b2);
\draw[thick,rounded corners=2pt] (gr3)--(1.7,2.2)--(b3);
\draw[thick,rounded corners=2pt] (b2)--(oc);
\end{tikzpicture}
\ \ \ \
\centering{=}\ \ \ \ \begin{tikzpicture}[cat,scale=1]
\node (i) at (0,-0.6) {};
\node (o) at (2,2.6) {};
\node (oc) at (1,2.6) {};
\node[black vertex] (b2) at (1,1){};
\ground{gr2}{0.7,1.4};
\draw[thick,rounded corners=2pt] (i) --(0,0)-- (b2)--(2,2) --(o);
\draw[thick,rounded corners=2pt] (gr2)--(0.7,1.2)--(b2);
\draw[thick,rounded corners=2pt] (b2)--(oc);
\end{tikzpicture}\]
i.e.~Alice, Bob and Eve share the same classical data.
\item Eve chooses a different  measurement than Alice and Bob in BB84:
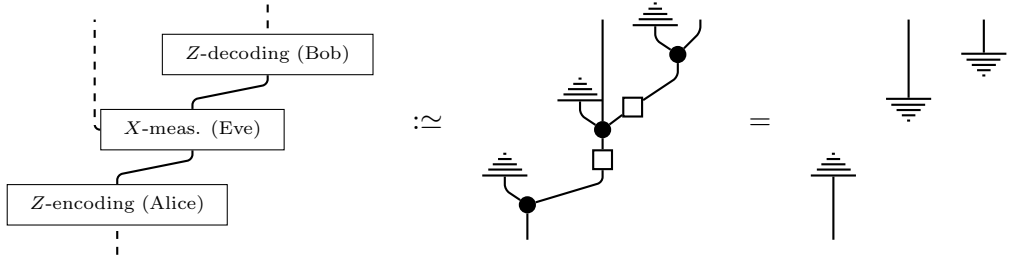
\[\begin{tikzpicture}[cat,scale=1]
\node (i) at (0,-0.8) {};
\node (o) at (2,2.8) {};
\node (oc) at (-0.3,2.6) {};
\node[box vertex] (m1) at (0,0){\tiny \begin{tabular}{c}$Z$-encoding (Alice)\end{tabular}};
\node[box vertex] (m2) at (1,1){\tiny \begin{tabular}{c}$X$-meas. (Eve)\end{tabular}};
\node[box vertex] (m3) at (2,2){\tiny \begin{tabular}{c}$Z$-decoding (Bob)\end{tabular}};
\draw[thick, dashed,rounded corners=2pt] (i) -- (m1);
\draw[thick, dashed,rounded corners=2pt] (o) -- (m3);
\draw[thick,rounded corners=2pt] (m1)   --(0,0.4)--(1,0.6)-- (m2) --(1,1.4)--(2,1.6)--(m3);
\draw[thick, dashed,rounded corners=2pt] (m2)--(-0.3,1)--(oc);
\end{tikzpicture}\ \ \ \ 
\centering{:\simeq}\ \ \ \ \begin{tikzpicture}[cat,scale=1]
\node (i) at (0,-0.6) {};
\node (o) at (2.3,2.6) {};
\node (oc) at (1,2.6) {};
\node[black vertex] (b1) at (0,0){};
\node[black vertex] (b2) at (1,1){};
\node[black vertex] (b3) at (2,2){};
\ground{gr1}{-0.3,0.4};
\ground{gr2}{0.7,1.4};
\ground{gr3}{1.7,2.4};
\node[h vertex](h1) at (1,0.6) {};
\node[h vertex](h2) at (1.4,1.3) {};
\draw[thick,rounded corners=2pt] (i) -- (b1) --(1,0.3)--(h1) -- (b2)--(h2) -- (2,1.7)-- (b3) -- (2.3,2.3)--(o);
\draw[thick,rounded corners=2pt] (gr1)--(-0.3,0.2)--(b1);
\draw[thick,rounded corners=2pt] (gr2)--(0.7,1.2)--(b2);
\draw[thick,rounded corners=2pt] (gr3)--(1.7,2.2)--(b3);
\draw[thick,rounded corners=2pt] (b2)--(oc);
\end{tikzpicture}
\ \ \ \ 
\centering{=}\ \ \ \ \begin{tikzpicture}[cat,scale=1]
\node (i) at (0,-0.6) {};
\node (o) at (2,2.6) {};
\node (oc) at (1,2.6) {};
\ground{gr1}{0,0.4};
\maxmixed{gr2}{1,1.4};
\maxmixed{gr3}{2,2};
\draw[thick,rounded corners=2pt] (i) --(gr1);
\draw[thick,rounded corners=2pt] (o) --(gr3);
\draw[thick,rounded corners=2pt] (gr2)--(oc);
\end{tikzpicture}\]
i.e.~Alice's,  Bob's and Eve's data is not correlated.
\eit
\end{prop}
 
\section{Connection to Selinger's CPM-construction}\label{sec:Selinger}

Here we briefly describe the connection between Definition \ref{def:environment} and the CPM-comstruction \cite{Selinger}, which was established in \cite{SelingerAxiom}.   Given any dagger compact category ${\bf C}$, we define a new category $CPM({\bf C})$ which has the same objects as ${\bf C}$, and a morphisms of type $A\to B$ in $CPM({\bf C})$ is a morphism of type $A\otimes A\to B\otimes B$ in ${\bf C}$ of the shape:
\beq\label{CPmap}
\begin{tikzpicture}[cat,scale=0.8]
\node (in1) at (0,3.5) {$B$};
\node (out1a) at (0.5,-2) {$A$};
\node (inn) at (0.8,3.5) {$B$};
\node (outna) at (2.5,-2) {$A$};
\node [morph] (f) at (0.5,-0.7) {$\ \ f\ \ $};
\node [dagger morph] (fd) at (2,1.8) {$\ \ f\ \ $};
\draw[thick,rounded corners=3pt]  (f) -- (out1a);
\draw[thick,rounded corners=3pt]  (fd) -- (2,2.6) -- (3.5,2.6) -- (3.5,0.5)--(2.5,-0.2) -- (outna);
\draw[thick] (0,-0.28) -- (in1);
\draw[thick,rounded corners=3pt] (1.1,-0.28) --(1.1,0.07)-- (2.5,1) -- (2.5,1.35);
\draw[thick,rounded corners=3pt] (inn)-- (0.8,1) -- (1.5,1) -- (1.5,1.35);
\end{tikzpicture}
\eeq
where $f:A\to B\otimes C$ is any morphism in ${\bf C}$.  Then, $CPM({\bf C})$ is again a dagger compact category, and as already mentioned in Example \ref{Ex:CPM}, if we set ${\bf C}:={\bf FHilb}$ then the morphisms of the form (\ref{CPmap}) are exactly completely positive maps. 

\begin{defi}   
An \em environment with  purification \em for $(\Cpure, {\bf C})$  is an environment as in Definition \ref{def:environment} for which we in addition have that, denoting morphisms in $\Cpure$ and more general morphisms in $\C$ respectively as
\[
\begin{tikzpicture}[cat,scale=0.8]
\node (in) at (0,-1) {$A$};
\node [morph] (f) at (0,0) {$f$};
\node (out) at (0,1) {$B$};
\draw[thick] (in) -- (f);
\draw[thick] (out) -- (f);
\end{tikzpicture}
\qquad\qquad\qquad\qquad
\begin{tikzpicture}[cat,scale=0.8]
\node (in) at (0,-1) {$A$};
\node [super box] (sp) at (0,0) {$F$};
\node (out) at (0,1) {$B$};
\draw[thick] (in) -- (sp);
\draw[thick] (out) -- (sp);
\end{tikzpicture}\ ,
\]
that for all $A, B\in |\C|, F\in\C(A,B)$ there exists $f\in \Cpure(A,B\otimes C)$ such that
\[
\begin{tikzpicture}[cat,scale=0.8]
\node (in) at (0,-1) {$A$};
\node [super box] (sp) at (0,0) {$F$};
\node (out) at (0,1) {$B$};
\draw[thick] (in) -- (sp);
\draw[thick] (out) -- (sp);
\end{tikzpicture}\ \ \ \ 
\centering{=}\ \ \ \ 
\begin{tikzpicture}[cat,scale=0.8]
\node (in) at (0.5,-1) {$A$};
\node [morph] (f) at (0.5,0) {$~~~f~~~$};
\node (out) at (0,1) {$B$};
\draw[thick] (in) -- (f);
\draw[thick] (out) -- (0,0.43);
\ground{gr}{1,0.8}
\draw[thick] (1,0.43) -- (gr);
\end{tikzpicture}\ .
\]
\end{defi}

\begin{thm} {\bf\cite{SelingerAxiom}}
$CPM(\Cpure)=\C$.
\end{thm}

As already mentioned in Example \ref{Ex:CPM},  the converse statement, that for any dagger compact category $\C$ the category  $CPM(\C)$ provides an environment with purification also holds, up to a minor and physically justified assumption related to the fact that vectors which are equal up to a complex phase represent the same state in quantum theory.  Concretely, this axiom states the for all pure elements $\psi, \psi':\II\to A$ we have:
\[ 
\psi\circ \psi^\dagger= \psi'\circ \psi'^\dagger \ \Rightarrow \ \psi= \psi'\ .
\]
This equation follows from Eq.~(\ref{eq:environment2}) when setting $f:=\psi^\dagger$ and $g:=\psi'^\dagger$.

\begin{rem}
The power of  purification as an axiom for quantum theory has recently been exploited in \cite{DAriano1,DAriano2}, although there, the authors also require certain uniqueness properties.
\end{rem}

\section{Conclusion}

An axiomatization of the concept of environment resulted in a very simple comprehensive graphical calculus, which in particular enables one to reason about classical-quantum interaction in quantum informatic protocols.

Several operationally distinct concepts turn out to have the same semantics within the graphical language (e.g.~classical channel, measurement, preparation as in BB84).  Consequently, all that one structurally truly needs are Propositions \ref{prop:idempotence} and \ref{pr:compunbchannels} on composition of classical channels and pure classical elements.

The examples given here are simple but representative.  This work and the earlier contributions on which we relied together successfully addresses  a challenge for the categorical quantum mechanics research program which was set at the very beginning: to have a very simple graphical description of  all basic quantum informatic protocols, in particular including classical-quantum interaction.

The new graphical element `environment' and the interaction rules for classical channels can now be integrated in the {\tt quantomatic} software, so that it can now be used to (semi-)automate reasoning about full-blown quantum informatic protocols, including classical-quantum interaction.

Here we only considered two complementary observables, and no phase data.  We meanwhile also have several graphical calculi that are universal for quantum computing \cite{CD2,CK}.  The next step of this research strand would be to extend the graphical calculus presented here to these calculi, which include, for example, phases and $W$-states.

This work could also be advanced in the direction of quantum information theory.  In particular, one may want to study whether it would be possible to obtain a diagrammatic account on quantum informatic quantities.  Some examples of diagrammatic quantum informatic quantities are in \cite{SelingerAxiom}.

\section*{Acknowledgements}

Work supported by EP/D072786/1, ONR  N00014-09-1-0248 and EU FP6 STREP QICS.  John Baez,  Aleks Kissinger, Prakash Panangaden, Johan Paulsson,  Jamie Vicary, the CSL'10  referees and the LMCS referees provided useful feedback on an earlier versions.


\end{document}

%% file: pics.tex
\def\state{\raisebox{2mm}{$\begin{tikzpicture}[cat,scale=0.8]
\node (in) at (0,1) {};
\node [black vertex] (gdot) at (0,0.4) {};
\draw[thick] (gdot) -- (in);
\end{tikzpicture}$}}

\def\effect{\raisebox{0mm}{$\begin{tikzpicture}[cat,scale=0.8]
\node (out) at (0,-1) {};
\node [black vertex] (gdot2) at (0,-0.4) {};
\draw[thick] (gdot2) -- (out);
\end{tikzpicture}$}}

\def\statebis{\raisebox{2mm}{$\begin{tikzpicture}[cat,scale=0.8]
\node (out2) at (0,1.4) {};
\node[black vertex] (gdot) at (0,0) {};
\node[h vertex] (h2) at (0,0.7) {};
\draw[thick] (out2)-- (h2) -- (gdot);
\end{tikzpicture}$}}

\def\effectbis{\raisebox{0mm}{$\begin{tikzpicture}[cat,scale=0.8]
\node (inc) at (0,-1.4) {};
\node[black vertex] (gdot) at (0,0) {};
\node[h vertex] (h1) at (0,-0.7) {};
\draw[thick]  (gdot) --(h1)-- (inc);
\end{tikzpicture}$}}

\def\cup{\raisebox{2mm}{$\begin{tikzpicture}[cat,scale=0.8]
\node (in1) at (-1,0.8) {};
\node (in2) at (1,0.8) {};
\draw[thick,rounded corners=8pt]  (in1) -- (-1,-0) -- (1,-0) -- (in2);
\end{tikzpicture}$}}

\def\cap{\raisebox{0mm}{$\begin{tikzpicture}[cat,scale=0.8]
\node (in1) at (-1,-0.8) {};
\node (in2) at (1,-0.8) {};
\draw[thick,rounded corners=8pt]  (in1) -- (-1,0) -- (1,0) -- (in2);
\end{tikzpicture}$}}

\def\had{\raisebox{0mm}{$\begin{tikzpicture}[cat,scale=0.8]
\node (out2) at (0,1.4) {};
\node(gdot) at (0,0) {};
\node[h vertex] (h2) at (0,0.7) {};
\draw[thick] (out2)-- (h2) -- (gdot);
\end{tikzpicture}$}}

\def\cnot{\raisebox{0mm}{$\begin{tikzpicture}[cat,scale=0.8]
\node (out) at (-1.4,1.4) {};
\node (out2) at (0,1.4) {};
\node[black vertex] (f) at (-1.4,0) {};
\node (inc) at (0,-1.4) {};
\node (inq) at (-1.4,-1.4) {};
\node[black vertex] (gdot) at (0,0) {};
\node[h vertex] (h) at (-0.7,0) {};
\node[h vertex] (h1) at (0,-0.7) {};
\node[h vertex] (h2) at (0,0.7) {};
\draw[thick,rounded corners=2pt] (gdot)  -- (h) -- (f);
\draw[thick] (inq)  -- (f) -- (out);
\draw[thick] (out2)-- (h2) -- (gdot) --(h1)-- (inc);
\end{tikzpicture}$}}

\def\cnotm{\raisebox{0mm}{$\left(\begin{array}{rrrr}
1 & 0 & 0 & 0\\
0 & 1 & 0 & 0\\
0 & 0 & 0 & 1\\
0 & 0 & 1 & 0
\end{array}\right)$}}

\def\trace{\raisebox{0mm}{\begin{tikzpicture}[cat,scale=0.8]
\node (in) at (0,-1) {};
\ground{gr}{0,0}
\draw[thick] (in) -- (gr);
\end{tikzpicture}}}

\def\decohere{\raisebox{0mm}{\begin{tikzpicture}[cat,scale=0.8]
\node (in) at (0,1.2) {};
\node (out) at (0,-0.8) {};
\ground{gr}{0.7,0.7}
\node [black vertex] (gdot) at (0,0) {};
\draw[thick,,rounded corners=2pt] (gdot) -- (0.7,0.5) -- (gr);
\draw[thick] (gdot) -- (in);
\draw[thick] (gdot) -- (out);
\end{tikzpicture}}}

\def\mix{\raisebox{0mm}{\begin{tikzpicture}[cat,scale=0.8]
\maxmixed{gr}{0,-0.6}
\draw[thick] (gr) -- (0,0.5);
\end{tikzpicture}}}
